# Performance of Exchange-Correlation Approximations to Density-Functional Theory for Rare-earth Oxides

*,1Mary Kathleen Caucci, 1Jacob T. Sivak, 2Saeed S.I. Almishal, 3Christina M. Rost, 2Ismaila Dabo, 2 Jon-Paul Maria, *,1,2,4,5Susan B. Sinnott

1Department of Chemistry, The Pennsylvania State University, University Park, PA 16802
2Department of Materials Science and Engineering, The Pennsylvania State University, University Park, PA 16802
3Department of Materials Science and Engineering, Virginia Polytechnic Institute and State University, Blacksburg, VA 24060
4Materials Research Institute, The Pennsylvania State University, University Park, PA, 16802
5Institute for Computational and Data Science, The Pennsylvania State University, University Park, PA, 16802
(Dated: April 13, 2024)

**Abstract**

Rare-earth oxides (REOs) are an important class of materials owing to their unique properties, including high ionic conductivities, large dielectric constants, and elevated melting temperatures, making them relevant to several technological applications such as catalysis, ionic conduction, and sensing. The ability to predict these properties at moderate computational cost is essential to guiding materials discovery and optimizing materials performance. Although density-functional theory (DFT) is the favored approach for predicting electronic and atomic structures, its accuracy is limited in describing strong electron correlation and localization inherent to REOs. The newly developed strongly constrained and appropriately normed (SCAN) meta-generalized-gradient approximations (meta-GGAs) promise improved accuracy in modeling these strongly correlated systems. We assess the performance of these meta-GGAs on binary REOs by comparing the numerical accuracy of thirteen exchange-correlation approximations in predicting structural, magnetic, and electronic properties. Hubbard $U$ corrections for self-interaction errors and spin-orbit coupling are systematically considered. Our comprehensive assessment offers insights into the physical properties and functional performance of REOs predicted by first-principles and provides valuable guidance for selecting optimal DFT functionals for exploring these materials.

## I. INTRODUCTION

First-principles computational approaches have historically struggled to accurately predict properties of rare-earth (RE) elements and compounds due to their highly correlated electronic structure with coexisting localized and itinerant (delocalized) states [1–4]. Capturing these complex electronic effects with first-principles methods is essential as many modern applications of RE systems are a direct result of the strongly interacting unpaired $d$- and $f$-electrons [5]. The 17 rare-earth elements consist of the lanthanide group (La—Lu) as well as Sc and Y which are included for their similar chemical behavior [6–8]. La is the first lanthanide element with no $4f$ electrons and an electronic configuration of $[Xe]4f^05d^16s^2$. Moving across the group adds an extra electron to the inner $4f$ shell rather than the outer $5d$ shell, resulting in a decrease in size. This systematic size reduction phenomenon, known as lanthanide contraction, is due to both incomplete screening of the $4f$ shells and relativistic effects, subsequently providing RE compounds with diversified physicochemical properties [7,9–12]. Rare-earth oxides (REOs) are receiving substantial attention for their potential use in many modern technological applications including solid ionic conductors, catalysis, efficient sensors, and neuromorphic computing [13–22]. REO materials characteristically possess mixed valences, high oxygen conductivities, low thermal conductivities, high melting temperatures, heavy fermionic behavior, and large dielectric constants [23–27]. The accurate modeling of these diverse electronic effects using *ab initio* approaches is fundamental for predicting structure-property relationships that drive material discovery and advance innovative applications of these systems. For example, rare-earth high-entropy oxides have recently emerged as candidates for thermochemical applications [28–30]. These materials undergo cation phase transitions that alter coordination environments, underscoring the need for a systemic study of their constituent oxide crystal structures as detailed in Methods Section B.

Density functional theory (DFT) [31,32] is the most widely utilized first-principles method for theoretically modeling materials at the electronic level because it provides a reasonable balance between accuracy and computational cost. Within the Kohn-Sham approach to DFT, the most complex electron interactions are collected into an exchange-correlation (XC) energy functional ($E_{xc}$). The exact functional form of the electron interactions contained in $E_{xc}$ is not known and therefore must be approximated. Hence, the accuracy of DFT predictions hinges upon the choice of XC functional used to model the electron-electron interactions [33,34]. Perdew and coworkers proposed an illustrative hierarchy, referred to as Jacob's ladder [35–37], that describes XC functionals in ascending accuracy by assigning $E_{xc}$ approximations to rungs on the ladder. As one moves up the ladder, the theoretical rigor increases, the XC approximations become more complex, and the energy functionals depend on additional information. The local density approximation (LDA) occupies the first rung of the ladder, depending only on the electron density $\rho$. Generalized gradient approximations (GGAs) occupy the next rung up where their $E_{xc}$ depends on both $\rho$ and its gradient $\nabla\rho$. Comprising the third rung of Jacob's ladder are the meta-GGAs which take the XC functional form of GGA and further add a dependency on the orbital kinetic energy density $\tau$. These three levels of theory are considered local (LDA) or semi-local (GGA, meta-GGA) functionals. Inexact treatment of electron exchange interactions underlying local and semi-local functionals leads to a deficiency of DFT that significantly influences its accuracy: delocalization, or self-interaction error (SIE) [38–40]. Moving up Jacob's ladder to higher levels of theory leads to greater accuracy and, in principle, improvement of the SIE, to varying degrees, at the expense of increased computational demand. This well-known error is particularly severe for systems with partially occupied *d* or *f* states, making the selection of $E_{xc}$ crucial to correctly describe these systems' electronic structure, magnetic ground state, thermodynamic properties, and relative energies [41–43]. To facilitate high-throughput predictions for solid-state materials, one of the most widely used $E_{xc}$ implementations within DFT is the Perdew-Burke-Ernzerhof (PBE) [44] parametrization of the GGA and its variants, such as PBEsol [45], as they present a reasonable compromise of accuracy and cost.

Various methods are used to mitigate the SIE within the DFT framework. One approach is to utilize non-local hybrid functionals, like HSE06 [46,47], which partially eliminates the SIE and improves the localization of *d* and *f* electrons by incorporating a fraction of exact Hartree-Fock exchange [43,48,49]. Seated on the fourth rung of the ladder, hybrid functionals provide greater accuracy in vibrational, electronic, thermodynamic, and structural properties, but is not widely utilized due to its high computational expense within periodic plane wave basis DFT calculations. Rather than plane waves, local basis sets have been implemented for REOs, but have been restricted to systems with small unit cells and simple magnetic structures [5,50]. A standard method for addressing the self-interaction error while avoiding the computational demands of hybrid functionals is to apply a Hubbard-type (*U*) parameter. The DFT+*U* [51,52] approach employs a Hubbard Hamiltonian for the Coulomb repulsion and exchange interaction to account for the strong on-site Coulomb repulsion amidst localized *4f* electrons [42,48,53]. The +*U* essentially acts as an on-site correction to reproduce the Coulomb interaction, thus serving as a penalty for delocalization. If the *4f* levels are partially filled, this potential is attractive and promotes the on-site *4f* electrons to localize. While there has been demonstrated success applied to REOs [24,33,43,54–56], the DFT+*U* method is not without its own shortcomings [42,48,57]. The description of metastable states can become problematic due to the initial orbital occupations, which can trap the structure in local minima [2,48]. In addition to the somewhat ad-hoc selection of the strength of the +*U* penalty, another, more serious, shortcoming is the limited transferability of a given DFT+*U* functional [42,57]. More recently, the strongly constrained and appropriately normed (SCAN) [58] and the restored regularized SCAN (r$^2$SCAN) [59] meta-GGA functionals are touted as a compromise of enhanced chemical accuracy with only a marginal cost increase from GGA [60]. The SCAN family of meta-GGA functionals has been shown to reduce the SIE for general materials and oxides, resulting in increased accuracy for property predictions [33,60–67]. The SCAN and r$^2$SCAN functionals may provide a solution to the limitations of expensive hybrid functionals or empirical corrections like +*U*, allowing accurate electronic-structure modeling for REOs within the same theoretical framework.

Characterizing REOs is challenging, as the high spatial localization of unpaired 4*f* electrons exhibit numerous possible on-site electronic and inter-site magnetic configurations. To capture these more complex electronic interactions, additional considerations are typically necessary to predict meaningful electronic properties. Typical DFT calculations are performed at the scalar relativistic level in which the scalar terms are treated with effective core potentials or pseudopotentials [68,69]. Treating valence electrons in this way fails for heavier atoms with larger nuclear charges, like the lanthanides, because the spin-orbit interactions become as strong as, or stronger, than the electron-electron repulsion term and may dominate the spin-spin or orbit-orbit interactions [70]. Consequently, the physical and chemical properties can be strongly influenced by these spin-orbit coupling (SOC) relativistic effects [12]. This SOC interaction can shift electronic levels, change the symmetry of electronic states, and describe the energetic splitting of atomic *p*, *d*, and *f* states [71,72]. Despite having significant influence, SOC is generally disregarded due to its increased computational cost and added complexity. However, in heavy-element systems, like REOs, where the SOC is strong, it can become necessary to include those effects in calculations to achieve even qualitatively accurate electronic descriptions, hence SOC effects are assessed as well.

Notwithstanding the recent promising reports of the SCAN and r$^2$SCAN functionals, their performance on binary REOs has not yet been investigated. We assess the ability of all aforementioned approximations to accurately predict formation energy, atomic structure, electronic structure, and magnetic character for the "light" REOs (*f* electron count < 7) including Y and Gd by comparing to experimentally measured values. Specifically, the performance, numerical accuracy, and computational efficiency of DFT, DFT+*U*, DFT+SOC, DFT+*U*+SOC are quantified across three different XC approximations (PBEsol, SCAN, or r$^2$SCAN) and two different pseudopotential parameterizations (4*f*-band and 4*f*-core). In total, thirteen different exchange-correlation approximations are considered. Our findings provide a guide to which DFT functionals are best suited for different properties and specific studies of REOs with respect to their computational expense.

## II. METHODS

### A. Computational Details

DFT calculations were performed using version 6.3.0 of the Vienna *ab initio* simulation package (VASP) [73,74] software utilizing the projector augmented-wave (PAW) method [75]. For the exchange-correlation functional, we employed the Perdew-Burke-Ernzerhof for solids (PBEsol) [45] generalized gradient approximation (GGA) functional, as well as the SCAN [58] or r$^2$SCAN [59] meta-GGA functionals. Unless otherwise indicated, calculations used pseudopotentials from the "PBE PAW datasets version 64" set released in April 2023. Although these pseudopotentials were developed for PBE functionals, they are commonly used with SCAN and r$^2$SCAN because there are no SCAN- or r$^2$SCAN-specific pseudopotentials available in VASP [62]. For all calculations, the plane wave kinetic energy cutoff was set to 700 eV with a convergence threshold of $10^{-6}$ eV for each self-consistent electronic calculation, and the ionic relaxations iterated until the atomic forces were less than 0.01 eV/Å. Γ-centered *k*-point grids were generated automatically by using KSPACING values of either 0.2 Å$^{-1}$ for oxides or 0.15 Å$^{-1}$ for metals. We used a two-step workflow which includes an initial GGA structure optimization followed by a structure optimization with the SCAN and r$^2$SCAN functionals using the GGA charge density as an initial guess in the meta-GGA optimization. More details on the two-step workflow approach can be found in Ref. [62]. For the density of states (DOS) calculations, the KSPACING was set in such a way as to increase each *k*-point grid by at least 2 along each direction, except for when SOC was used where the grid remained unchanged.

SOC calculations were initialized with the spin quantization axis along the z-direction (SAXIS = 0 0 1) and the moments were allowed to self-consistently solve. SOC is incorporated into calculations via the spin-orbit-coupled effective, two-component, Hamiltonian, and thus can be added to the Hamiltonian irrespective of the functional approximation [76,77]. Additional details and results can be found in the Supplemental Material [78]. SCAN with SOC was not considered due to slow convergence.

Core electrons were treated using the PAW [75] method with the following PBE PAW datasets [69]: Ce, Ce_3, Eu, Eu_3, Gd, Gd_3, La, Nd, Nd_3, O, Pr, Pr_3, Sm, Sm_3, and Y_sv—the notation _3 refers to the PAW treating the *4f* electrons as frozen-core electrons with the pseudopotential created for a valency of 3. The "standard" version treats the entire set of *f*-levels within the valence band. Any calculations using the _3 pseudopotentials will be referred to as the '*f*-core' or 'DFT-core' approach, whereas the '*f*-band' approach refers to using the potentials in which *f*-electrons are treated as valence.

TABLE I. Hubbard *U* values (in eV) of elements for respective XC-functionals

| Element | Functional | |
|---|---|---|
| | PBEsol | SCAN & r$^2$SCAN |
| O | 0 | 0 |
| Y | 0 | 0 |
| La | 3 | 1 |
| Ce | 6 | 4 |
| Pr | 4 | 2.5 |
| Nd | 2.5 | 1 |
| Sm | 7 | 5 |
| Eu | 7 | 5 |
| Gd | 7 | 3 |

We employed an onsite Hubbard-type (*U*) parameter to the *f*-states (*d*-states for La) through the simplified rotationally invariant framework developed by Dudarev *et. al.* [51]. *U* values were chosen in a semi-empirical way that when compared to experimental properties there is sufficient improvement to electronic predictions without significantly worsening the structural predictions. For simplicity, a Hubbard *U* was not added to the oxygen sites, but doing so has shown to yield better electronic structure for transition metal oxides [79]. Ancillary testing was performed for a variety of Coulomb values, where Fig. S3 exemplarily shows the calculated lattice parameters and band gaps for $CeO_2$ and $C-Ce_2O_3$ as a function of increasing *U* values. A single 'average' *U* parameter was established for each element (Table I), i.e. a *U* value of 4 eV is used for both $PrO_2$ and $Pr_2O_3$ in PBEsol. It should be noted that the values selected are likely not the optimal values across all properties but are chosen to be representative of a sufficient correction for REOs. Previous work has validated fitting properties to obtain an average *U* value that accounts for changes in formal oxidation state as well as different structures with diverse bonding [49,61]. Determining *U* values using a linear response approach was not considered since VASP currently does not implement a non-self-consistent field calculation for meta-GGA functionals which is necessary for linear response *U* [80]. To avoid convergence to metastable electronic states, the ramping method of Meredig *et. al.* was implemented [81]. Since the SIE of $f$-electrons is not present when treating the $4f$ electrons as core, the $f$-core calculations do not receive Hubbard $U$ corrections. In total, there are thirteen exchange-correlation approximations: PBEsol, PBEsol+$U$, PBEsol+SOC, PBEsol+$U$+SOC, SCAN, SCAN+$U$, r$^2$SCAN, r$^2$SCAN+$U$, r$^2$SCAN+SOC, r$^2$SCAN+$U$+SOC, PBEsol-core, SCAN-core, r$^2$SCAN-core.

Bader analysis was used to quantify the atomic charges and site-projected magnetic moments. Bader atomic volumes were computed by the program of Henkelman and co-workers using the total electron density [82–84]; then the valence and spin densities were integrated over these volumes to determine the Bader atomic charges and atomic spin moments (ASMs), respectively [85]. For our analysis, we used the approximate all-electron charge density (AECCAR + CHGCAR from VASP). The Bader net ionic charges are determined via the subtraction of the Bader projected valence charges from the valence electron count in the PAW potentials.

### B. Crystal structures

Figure 1 illustrates the four crystalline structures that binary REOs typically present at ambient conditions. All rare-earth elements form sesquioxides ($R_2O_3$, R = rare-earth element) which commonly display three different polymorphs below 2,000 °C [14,26,27,86]: A-type hexagonal ($P\bar{3}m1$, No. 164), B-type monoclinic ($C2/m$, No. 12), and C-type cubic ($Ia\bar{3}$, No. 206), commonly known as bixbyite. The A-type polymorph is favored by the earlier lanthanides (La-Nd), Y favors the bixbyite structure, and the smaller lanthanides (Sm-Gd) can form either the B- or C-type structures. In their most oxidized state, Ce and Pr take the cubic fluorite structure ($Fm\bar{3}m$, No. 225). Bixbyite is a low-symmetry variant of fluorite in which one-fourth of the oxygen sites are vacant in an ordered way. Throughout the text, sesquioxide structures are distinguished by labeling A-$R_2O_3$, B-$R_2O_3$, or C-$R_2O_3$. We note that $Pm_2O_3$ is not considered as it is highly unstable, radioactive, and not found in nature [87,88].

Table II summarizes the rare-earth elements' common valences and crystal structures. Primitive cells were used for all sesquioxide polymorphs, while the conventional 12-atom unit cell was used for the fluorite structure. The trigonal A-type phase has a primitive unit cell containing one formula unit (five atoms). The B-type conventional cell has 30 atoms with a cell volume that is 2x larger than its 15-atom primitive cell counterpart. The bixbyite C-type structure has 80 atoms in its conventional cell, whereas its rhombohedral primitive cell contains 40 atoms (8 formula units). For A-type sesquioxides and PrO$_2$, both ferromagnetic (FM) and antiferromagnetic (AFM) magnetic orderings were considered with AFM ordering used for all property predictions. The A-type hexagonal $P\bar{3}m1$ contains two cations per unit cell leading to a broken-symmetry antiferromagnetic spin configuration. The AFM ordering for the A-type sesquioxide has one spin up and one spin down per unit cell. The A-type-AFM ordering was used for PrO$_2$ [89–91]. All other REOs and metallic REs were initialized in FM only.

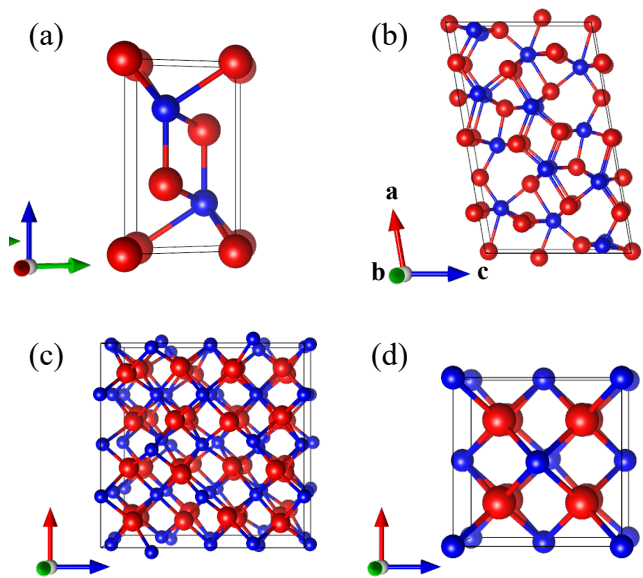

FIG. 1. Conventional unit cell structures of the $R_2O_3$ sesquioxides for (a) hexagonal A-type, (b) monoclinic B-type, (c) the bixbyite C-type, and (d) the $RO_2$ fluorite. R=blue and O=red.

TABLE II. Common valence, valence electron configuration, and considered crystalline structures for the light RE elements with the inclusion of Y and Gd.

| Element Symbol | Common valence | Valence electron configuration | | | | Crystalline structure |
|---|---|---|---|---|---|---|
| | | Atomic | +2 | +3 | +4 | |
| Y  | +3      | $4d^1 5s^2$      |        | $4d^0$ |        | C-type |
| La | +3      | $4f^0 5d^1 6s^2$ |        | $4f^0$ |        | A-type |
| Ce | +3, +4  | $4f^1 5d^1 6s^2$ |        | $4f^1$ | $4f^0$ | A-type, Fluorite |
| Pr | +3, +4  | $4f^3 6s^2$      |        | $4f^2$ | $4f^1$ | A-type, Fluorite |
| Nd | +3      | $4f^4 6s^2$      |        | $4f^3$ |        | A-type |
| Sm | +2[a], +3 | $4f^6 6s^2$    | $4f^6$ | $4f^5$ |        | B-type & C-type |
| Eu | +2[a], +3 | $4f^7 6s^2$    | $4f^7$ | $4f^6$ |        | B-type & C-type |
| Gd | +3      | $4f^7 5d^1 6s^2$ |        | $4f^7$ |        | B-type & C-type |

[a]The 2+ oxidation state can also be found for both Sm and Eu in rock salt structures [14,26,27].

## III. RESULTS

### A. Structural properties: crystal structures and lattice volumes

The accuracy of predicted crystal structures is evaluated by the computed lattice parameters and the relaxed lattice volume in comparison to their respective experimentally determined values [26,27,92–95,95–106]. For all thirteen exchange-correlation approximations considered (including *f*-core potentials), the error for every lattice parameter is approximately less than 1%, except for the A-type REOs when using PBEsol (~2%). Figure 2 illustrates the mean relative errors (MREs) and mean absolute errors (MAEs) for predicted lattice volumes per atom for DFT relative to experiment for all thirteen XC approximations considered. The sign of the MRE indicates whether calculated values are underpredicted (negative) or overpredicted (positive) compared to experiment. For the binary REOs, all but PBEsol, SCAN, and PBEsol-core yield a MRE of < ±1% on the total volume; r$^2$SCAN has the lowest MAE (0.113 Å$^3$/atom), which is followed by SCAN+$U$ (0.156 Å$^3$/atom). Overall, the meta-GGA functionals show a clear improvement in predicting binary REO unit cell volumes in comparison to PBEsol which has the highest MAE of 0.525 Å$^3$/atom. The meta-GGA functionals' ability for improved volume is likely a result of their smaller SIEs. As observed in Fig. 2(b), all three functionals systematically underestimate the unit cell volumes, where r$^2$SCAN consistently predicts larger volumes than PBEsol or SCAN. Unlike SCAN and r$^2$SCAN, the PBEsol approximations do not maintain the decreasing periodic trends of lanthanide contraction for A-Pr$_2$O$_3$, A-Nd$_2$O$_3$, B-Sm$_2$O$_3$, and C-Sm$_2$O$_3$ [Fig. S5]. When a Hubbard $U$ correction is applied, the MAE for PBEsol+$U$ (0.222 Å$^3$/atom) is reduced by ~60% compared to PBEsol, and the decreasing periodic trend is restored. As for SCAN and r$^2$SCAN, Hubbard $U$ corrections result in lattice volumes going from slight underprediction to slight overprediction. Moreover, while the MAE worsens for r$^2$SCAN+$U$ to 0.169 Å$^3$/atom, there is an error improvement of ~15% for SCAN+$U$ [Fig. 2(a)]. As will be discussed in Section IV, this is a consequence of the 'average' $U$ value applied in our study that allows for proper electronic structure predictions. The consistent increase in DFT+$U$ lattice volumes reduces SCAN's MRE while increasing the MRE of r$^2$SCAN, which has accurate structural predictions without any corrections. Including the rVV10 van der Waals correction slightly improves the r$^2$SCAN+$U$ MAE but worsens the SCAN+$U$ MAE [Fig. S6], in agreement with previous observations by Kothakonda *et. al.* [64]. For PBEsol, SCAN, and r$^2$SCAN, the $f$-core potentials perform comparably to one another with each having a respective MAE (MRE) of 0.346 Å$^3$/atom (-1.30%), 0.349 Å$^3$/atom (-0.14%), and 0.304 Å$^3$/atom (0.83%). It should be noted that when comparisons between predictions of $f$-core to $f$-band are made, the $f$-band results only include those that have $f$-core options (i.e., not including Y$_2$O$_3$, La$_2$O$_3$, PrO$_2$, CeO$_2$). The $f$-core approach does not apply to the fluorite oxides CeO$_2$ and PrO$_2$ because the '_3' pseudopotentials cannot describe their 4+ valence resulting in large errors in volume, energetic, and electronic predictions [Tables S1, S4, S6]. We also note that including SOC could either improve or worsen lattice volume predictions [Table S1].

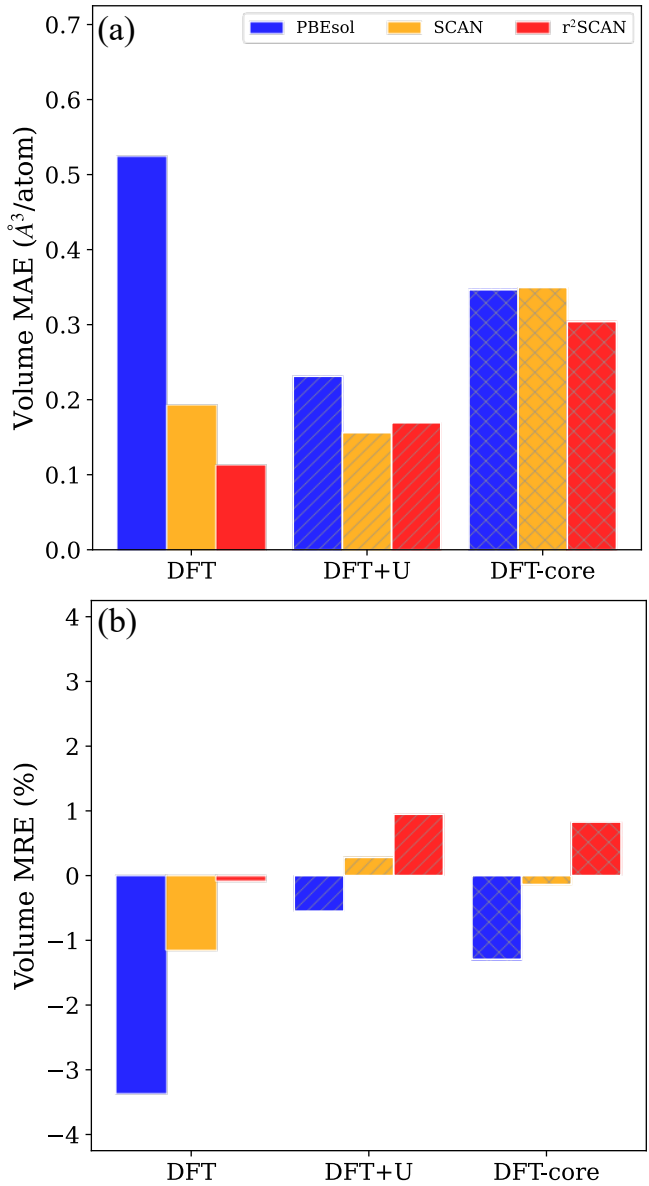

FIG. 2. The (a) mean absolute error, MAE, and (b) mean relative error, MRE for DFT, DFT+$U$, and DFT-core lattice volume predictions compared to experiment.

### B. Electronic structure

#### 1. Band gaps

REOs exhibit various metastable states due to their unpaired $f$ electrons' high localization, resulting in distinct electronic behaviors such as metallic or insulating characteristics [5]. An accurate description of these strongly localized $f$ electrons is crucial for reliable descriptions of REOs' electronic properties [107]. Figure 3 schematically depicts the relevant band gap structures of rare-earth sesquioxides (R$_2$O$_3$), highlighting the conduction band, valence band, and the narrow occupied ($f^n$) and empty ($f^{n+1}$) 4$f$ bands. The positioning of the $f$ states relative to the band edges as well as the splitting of the $f$-bands into sub-bands are distinguishing features of rare-earth sesquioxide (RES) gap structures. Assessing the partial density of states (pDOS) helps determine the functionals' ability to predict the relative positions of the $f$-bands, which significantly influence the mechanism of electronic conduction. It is important to note that, within the Kohn-Sham scheme of ground-state theory, DFT is not expected to precisely reproduce measured experimental band gaps [108]. Yet, ground-state functionals like those considered here, are expected to be well equipped to predict the ground-state band gaps and electronic structure of these REOs.

Examining individual REOs, the gap structure varies based on the occupation of the $f$-bands. At the beginning of the rare-earth series, there are no localized bands within the $s$-$d$ gaps. Besides La$_2$O$_3$ (Y$_2$O$_3$), which has an empty $f$-band ($d$-band), all other light RE sesquioxides have a localized $f$-state between the O-2$p$ and RE $d$-band. The empty 4$f$ band results in 4$f$ levels being within the conduction band and produces a fundamental transition occurring from the O-2$p$ valence to the RE-5$d$ conduction band [Fig. 3(a)]. This gap structure is easily reproduced by DFT [Fig. 3(d)-(f)]. As the number of 4*f* electrons increases across the lanthanide series, the electron density of the occupied $f^n$ increases, the energy of the 4*f* state decreases, and the position of the occupied $f^n$ begins to fall below the conduction band [109].

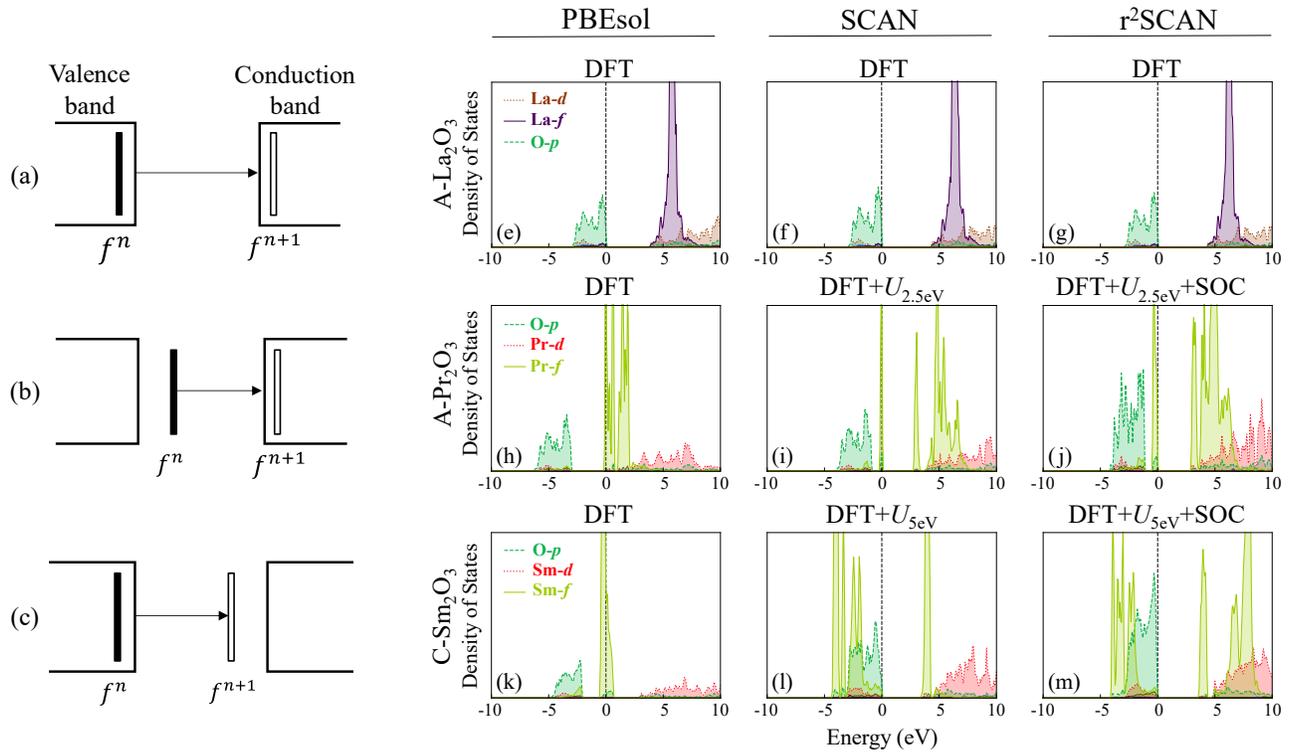

FIG 3. (a) – (c) Diagrams illustrating the energy band gap of RE sesquioxides. The $f^n$ and $f^{n+1}$ represent the occupied and unoccupied *f* bands, respectively, and can be situated on either side of the band edges. The (a) band gap is relevant to $Y_2O_3$, $La_2O_3$, and $Gd_2O_3$. The (b) band gap is relevant to $Ce_2O_3$, $Pr_2O_3$, and $Nd_2O_3$. The (c) band gap is relevant to $Sm_2O_3$ and $Eu_2O_3$. (e) – (m) Electronic pDOS of the spin-up channel for (e) – (g) A-$La_2O_3$, (h) – (j) A-$Pr_2O_3$, and (k) – (m) C-$Sm_2O_3$ using either DFT, DFT+$U$, or DFT+$U$+SOC. Fermi level is set to 0 and is indicated with a black dotted line.

The process of $f^n$ and $f^{n+1}$ continuously decreasing with the increase of 4*f* occupation repeats itself starting with $Gd_2O_3$, where the fully half-filled and empty *f*-bands reside within the valence and conduction bands, respectively [Fig. 3(a)]. The more iterant *f*-bands of $Gd_2O_3$ allow even PBEsol to capture the appropriate insulating character. Figure 4 compares the predicted fundamental band gaps to experimental measurements [27,94,98,102,105,110–115], revealing periodic variations in band gap values and characteristics. Although placing the *f*-electrons in the core produces the proper insulating electronic behavior for REOs, the model fails to maintain qualitatively correct band gap characteristics since the localized *f*-bands are not preserved [Fig. 4(c)]. In the band gap structure illustrated in Fig. 3(b), the transition occurring from the localized $f^n$ state to the conduction band is pertinent to $Ce_2O_3$, $Pr_2O_3$, and $Nd_2O_3$. For C-$Ce_2O_3$, DFT-induced delocalization leads to improperly predicting metallic character for all three DFT approximations [Fig. 4(a)], and a similar non-insulating outcome is observed for A-$Ce_2O_3$ and A-$Pr_2O_3$ with PBEsol. The application of Hubbard $U$ localizes the partially filled *f*-orbital, shifting the Fermi energy, and expanding the band gaps for these oxides. Whereas SCAN and $r^2$SCAN can produce an insulating state for all A-type oxides without a +$U$ correction, albeit with significantly underestimated fundamental gaps. However, despite producing an insulating state for A-$Pr_2O_3$, at either the DFT (SCAN, $r^2$SCAN) or DFT+$U$ (PBEsol) levels, the band edge orbital characters are wrongly assigned and require SOC to properly amend the 4*f* states [Fig. S10].

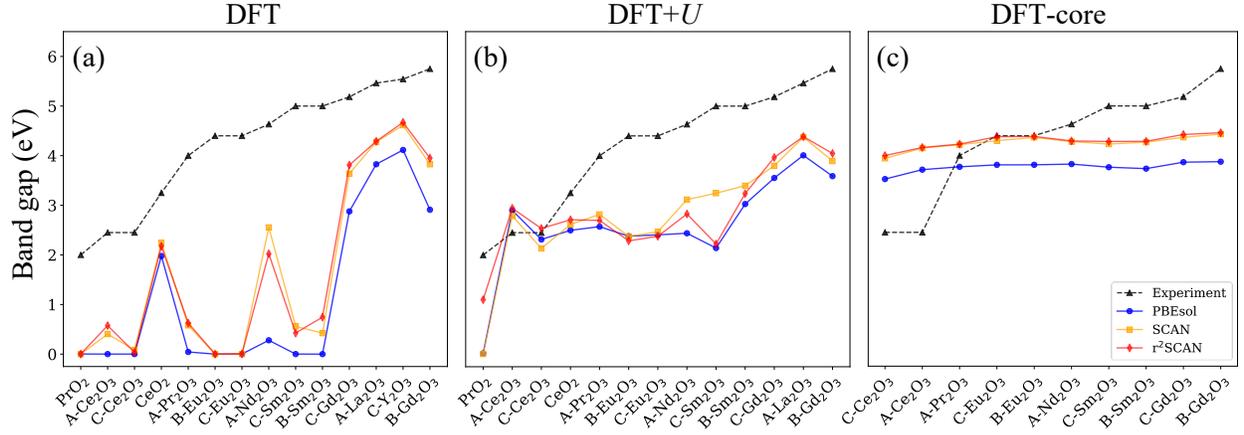

FIG 4. Comparison of experimental band gaps to predicted minimum electronic band gaps as obtained using PBEsol (blue circles), SCAN (yellow squares), and r²SCAN (red diamonds) in (a) DFT, (b) DFT+$U$, and (c) DFT-core. For visual clarity, data is sorted by experimental values and lines are to guide the eye.

Continuing across the row to $Sm_2O_3$ and $Eu_2O_3$, the occupied $f^n$ band becomes hybridized with the valence band, while the empty $f^{n+1}$ localizes within the gap [Fig. 3(c)]. PBEsol performs inadequately with the highly correlated oxides of Sm and Eu. The meta-GGAs struggle to realize an insulating state for $Eu_2O_3$ without +$U$, but they do produce an insulating state for $Sm_2O_3$. Inspecting the predicted band gap structure for B-$Sm_2O_3$ and C-$Sm_2O_3$ reveals that, except for SCAN+$U$, the insulating states predicted for C-$Sm_2O_3$ are unphysical and do not align with Fig. 3(c). Appropriate electronic description and band gap character for C-$Sm_2O_3$ are achieved only when SOC is included [Fig. 3(m)], while B-$Sm_2O_3$ does not require SOC [Fig. S12]. Similarly, the fluorite-structured $PrO_2$ and A-$Pr_2O_3$ require SOC for qualitatively accurate band gap structures [116], except for r²SCAN+$U$, which doesn't need SOC for $PrO_2$ [Fig. S11].

Further, the choice of pseudopotential and XC approximation directly influences the level of theory necessary to predict qualitatively accurate band gap structures and density of states. Before the recent release of VASP's 'version 64' PAW dataset, the 52 and 54 PAW dataset versions had been extensively utilized, but the newer version 64 PAW dataset includes updated $f$-band pseudopotentials for Pr, Nd, Sm, Eu, and Gd. For $Sm_2O_3$, the PAW datasets (versions 52, 54, and 64) combined with PBEsol and PBEsol+$U$ incorrectly predict a metallic state, in contrast to the approximately 4.75eV experimental band gap [117]. Correcting this requires including SOC and +$U$, particularly with the 52 [Fig. S14] or 64 versions [Fig. S13]. Interestingly, with the 54 version, only PBEsol+$U$ is necessary to produce the proper band gap characterization and high localization of $f$-states. When using meta-GGAs, SOC remains required with the 54 version, but the 64 version paired with SCAN can predict the electronic structure at the DFT+$U$ level, provided a sufficiently high $U$ value [Fig. S13]. For $PrO_2$, all XC functionals require SOC in the 54 version. However, in the 64 version, r²SCAN accurately captures the band gap character at the DFT+$U$ level. Thus, both SOC and +$U$ are essential for addressing strong relativistic effects and the high localization of $f$-states, especially in C-$Sm_2O_3$ and fluorite-structured $PrO_2$. While using the newest 64 version is highly recommended, this illustrates how sensitive the choice of pseudopotential impacts the accuracy of predicted band gap structures in REOs.

Collectively, the electronic behavior of $CeO_2$, C-$Y_2O_3$, A-$La_2O_3$, A-$Nd_2O_3$, B-$Gd_2O_3$, and C-$Gd_2O_3$ is effectively captured by PBEsol, SCAN, and r²SCAN. Unlike PBEsol, the meta-GGA functionals exhibit the ability to predict insulating behavior for A-$Ce_2O_3$ and B-$Sm_2O_3$ without empirical corrections. However, all three functionals struggle with C-$Ce_2O_3$, B-$Eu_2O_3$, and C-$Eu_2O_3$, where the lack of insulating behavior necessitates the incorporation of +$U$ to achieve adequate band gap structures. For PrO$_2$, both PBEsol and SCAN require the joint application of +$U$ and SOC for accurate electronic representation, whereas r$^2$SCAN attains proper behavior at the DFT+$U$ level. Despite yielding insulating states for A-Pr$_2$O$_3$ and C-Sm$_2$O$_3$ without +$U$, the meta-GGAs (and PBEsol+$U$) require the inclusion of SOC due to the incorrect band edge character from erroneous localization of 4$f$ states too close to the Fermi level. These nuanced findings underscore the necessity of tailoring computational approaches and emphasize the interplay between various functionals, corrections, and oxide systems for a comprehensive understanding of their electronic properties.

### 2. *Electron localization*

Bader charges are calculated to assign the total charge associated with each atom, indicating the extent of electron localization achieved by an approximation. The atomic charges are subsequently taken as the average of the Bader charges on the rare-earth cations in the oxide compound. Figure 5 depicts the total average Bader charge for the REOs using different density functional approximations, which can be correlated to formal oxidation states. The Bader charge values are lower than the formal oxidation states of 3+ or 4+ due to electrons not sitting fully on atom centers, implying the presence of some degree of covalent bonding character with O-2$p$ valence electrons. These lower Bader charges are consistent with previous work [43,85]. In all cases, PBEsol Bader charges are consistently lower than the values predicted by its meta-GGA counterparts, emphasizing the enhanced ability of meta-GGAs to mitigate the delocalization inherent in GGAs like PBEsol.

Expectantly, the introduction of a Hubbard $U$ term further increases the atomic charges, to varying extents, across all three functionals, as it enables greater electron localization on the rare-earth $d$/$f$ orbitals. A-Nd$_2$O$_3$, B-Gd$_2$O$_3$, and C-Gd$_2$O$_3$ have the smallest increases in atomic charges when going from DFT to DFT+$U$. This trend is also observed in the cases of A-Pr$_2$O$_3$ and PrO$_2$ for r$^2$SCAN and PBEsol, respectively. Although the atomic charges from PBEsol+$U$ remain lower than SCAN+$U$ and r$^2$SCAN+$U$, the difference in Bader charges between meta-GGA and GGA is lower within DFT+$U$ than in comparison to DFT [Fig. S15].

When assessing differences between $f$-band and $f$-core pseudopotentials, the average $f$-band atomic charges from DFT are lower than $f$-core, however, the $f$-band values from DFT+$U$ become comparable to those from their respective $f$-core approximations. The $f$-core pseudopotentials are optimized for the 3+ oxidation states and therefore have fewer delocalization errors from the absence of valence *f*-electrons. Thus, the lower average atomic charges from the DFT *f*-band are a result of the persistence of delocalization errors. The *f*-band average atomic charges from DFT+*U* become similar to the *f*-core values which reflects the mitigation of the delocalization through the +*U* correction. Incorporating SOC does not greatly influence the Bader charges or change the observed trend [Table S7].

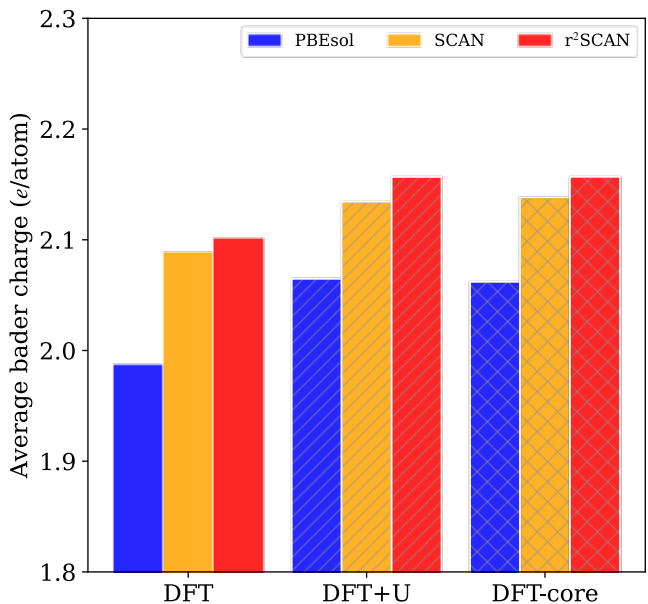

FIG 5. The average rare-earth Bader charge in $e$/atom for DFT, DFT+$U$, and DFT-core.

### 3. *Magnetic conformations and spin densities*

The partitioning of electron spin density among atoms gives rise to atomic spin moments (ASMs), which are important for understanding magnetic properties and allows one to quantify the type of magnetic state (ferromagnetic, antiferromagnetic, etc.) of a material [85]. The effective magnetic moments are taken as the average of the absolute atomic spin moment values of the rare-earth elements in the oxide compound. Since the *f*-electrons determine magnetic properties, freezing *f*-electrons in the core, as is done with *f*-core pseudopotentials, does not result in magnetic moments.

Figure 6 illustrates the difference in atomic spin moment $\Delta ASM$ between DFT+*U* and DFT for the REOs. The average magnetic moment predicted by PBEsol increases by 1.64% when Hubbard *U* is applied. The largest increase of more than 150% is made for A-$Ce_2O_3$, distantly followed by $PrO_2$ with an increase of about 32%. Comparatively, SCAN's average ASM increases by 0.27% while $r^2$SCAN's decreases by 0.52%. Unlike with Bader charges where a +*U* correction uniformly increases the atomic charges, there are variations in how DFT+*U* impacts the ASMs. From Fig. 6, we see that for $Sm_2O_3$ and $Eu_2O_3$ sesquioxides, the magnetic moments decrease with the introduction of +*U*. We also observe that for $PrO_2$ the +*U* correction led to a substantial decrease in ASM for $r^2$SCAN, and $r^2$SCAN+*U* had a 24% lower ASM than PBEsol+*U* and SCAN+*U*. Although SCAN and $r^2$SCAN correctly favor AFM ordering, only $r^2$SCAN+*U* appropriately produces an insulating AFM state as shown in Fig. 4(b). The dynamic Jahn-Teller distortion and strongly correlated nature of $PrO_2$ help explain the anomalous behavior seen in Fig. 6 [89,91,116,118–120]. Without SOC, typical DFT approaches can't capture PrO$_2$'s interesting properties like its insulating AFM ordering and its low effective magnetic moment 0.68μ$_B$ [89]. Even with the improvement that comes from including SOC, the predicted 1.14μ$_B$ magnetic moment from r$^2$SCAN+$U$ overshoots experimental estimates, possibly because we aren't accounting for vibronic effects [91].

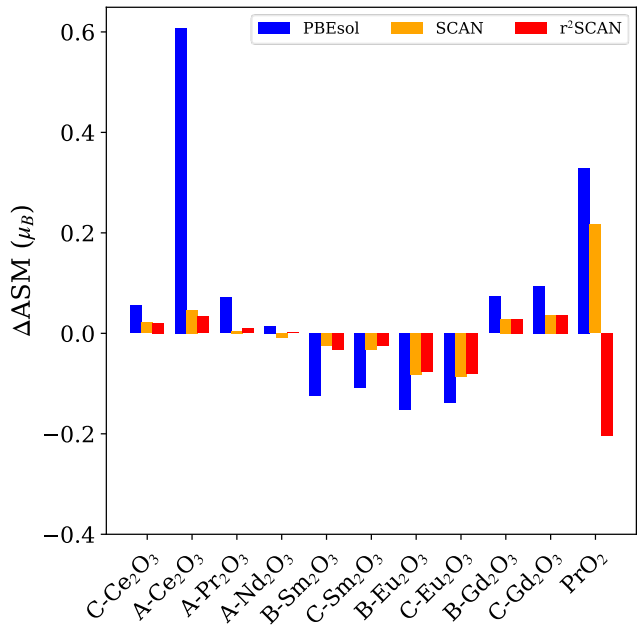

FIG 6. The difference of atomic spin moment $\Delta ASM$ = $ASM_{DFT+U} - ASM_{DFT}$ in $\mu_B$ between DFT+*U* and DFT. Values above zero indicate an increase in magnetic moment with +*U*, while values below zero indicate a decrease in ASM.

Without a +$U$ correction, PBEsol predicts lower effective magnetic moments in comparison to SCAN and r$^2$SCAN [Fig. S8], except for the Sm$_2$O$_3$ and Eu$_2$O$_3$ sesquioxides in which PBEsol has slightly larger magnetic moments than the meta-GGA functionals. When all REOs are considered, the average magnetic moment produced by SCAN and r$^2$SCAN is both ~2% higher than PBEsol and has mean relative differences of ~15%. These differences between the predictions made by PBEsol and the meta-GGA functionals lessen when Hubbard $U$ is included, such that the average mean relative difference between all three approximations is less than 1.0%. Yet within DFT+$U$, PBEsol still produces lower effective magnetic moments than the meta-GGA functionals except for PrO$_2$. Comparably for the metallic RE solids, the meta-GGAs predict larger effective magnetic moments than PBEsol [Fig. S3]. In agreement with prior observations, we find a similar tendency for SCAN and r$^2$SCAN to predict larger moments than other semi-local approximations [64,121]. Unlike previous work that found these meta-GGAs to produce worse experimental agreement than PBE due to over-magnetizing elemental transition metals, we find all three approximations similarly either over- or under-estimate experimental moments for elemental RE and REOs, see Sec. S1 in Supplemental Material [78]. The only instance in which PBEsol underpredicts ASM where the meta-GGAs overpredict is for metallic Pr and Gd.

To assess the topic of magnetic ordering, the predicted relative stability $\Delta E_{FM-AFM}$ (in meV/atom) of the ferromagnetic (FM) and antiferromagnetic (AFM) spin ordering is evaluated for the three magnetic A-type REOs and the fluorite-structured PrO$_2$. Table III presents $\Delta E_{FM-AFM}$ values for these four oxides using six different functionals. Experimentally, the AFM order is favored by the four oxides, however, PBEsol consistently predicts FM ordering as the more stable configuration, indicated by negative values. Table III highlights that a +$U$ correction is necessary for PBEsol to yield a more stable AFM ordering for the A-type sesquioxides. For PrO$_2$, both a Hubbard $U$ and SOC are needed for PBEsol to favor AFM configuration. SCAN properly describes a more favorable AFM configuration for all four oxides without a +$U$ correction; excluding Nd$_2$O$_3$ and Pr$_2$O$_3$, the relative stability of the AFM ordering increases upon applying a +$U$ correction. The same overall trend is realized for r$^2$SCAN, apart from Pr$_2$O$_3$, in which a stable AFM ground state is predicted going from r$^2$SCAN to r$^2$SCAN+$U$.

TABLE III. Relative stability $\Delta E_{FM–AFM} = E_{FM} − E_{AFM}$ (in meV/atom) of the ferromagnetic and antiferromagnetic spin configurations in Ce$_2$O$_3$, Pr$_2$O$_3$, Nd$_2$O$_3$, and PrO$_2$. Positive values indicate predicted AFM behavior, while negative values indicate predicted FM behavior.

|  | DFT | | | DFT+$U$ | | |
| --- | --- | --- | --- | --- | --- | --- |
| Compound | PBEsol | SCAN | r$^2$SCAN | PBEsol | SCAN | r$^2$SCAN |
| Ce$_2$O$_3$ | -20.6 | 23.8 | 54.0 | 513 | 401 | 431 |
| Pr$_2$O$_3$ | -46.9 | 0.489 | -4.26 | 17.4 | -0.0434 | 0.0318 |
| Nd$_2$O$_3$ | -14.3 | 28.2 | 34.1 | 26.4 | 9.45 | 0.768 |
| PrO$_2$ | -5.90 | 76.6 | 96.4 | -10.9 | 227 | 315 |

## C. Formation Energies

Rare-earth oxides, in particular the lanthanide oxides, exhibit exceptional thermal stability as a result of their electronic configuration and atomic size [14]. The thermodynamic stability of solid compounds is commonly described by the formation enthalpy $\Delta H_f$, a quantity extensively employed in literature, as opposed to the Gibbs free energy [122].

The general formation reaction equation of a RE binary oxide, $R_xO_y$, is written as

$$xR + \frac{y}{2}O_2 \rightarrow R_xO_y. \quad [1]$$

We calculate the enthalpy of formation, $\Delta H_f$, as the energy released when a compound is formed from the elemental constituents in their standard states [66]:

$$\Delta H_f^{298K} \approx \Delta E_f^{0K} = E - \sum_i x_i \mu_i \quad [2]$$

where $\Delta E_f^{0K}$ is the formation energy calculated by DFT, $E$ is the total energy of a compound containing $x_i$ atoms of element $i$, which has an elemental chemical potential of $\mu_i$ [60,62]. The chosen elemental reference states are specified in the Supplemental Material [Table S3]. Assuming the oxide's $PV$ term is negligible and zero-point fluctuations are disregarded, the formation enthalpy $\Delta H_f$ can be approximated from the DFT formation energy $\Delta E_f$, allowing $\Delta H_f \approx \Delta E_f$. To maintain a consistent, physically meaningful quantity, electronic energy differences are calculated using the same level of theory for both REOs and their elemental components within a given DFT approximation. For instance, DFT+$U$ calculations apply the same $U$ value to the $f$-orbitals in both the oxide and the metallic RE species to ensure consistent energy references. Similarly, the $f$-core approximation is applied uniformly to both oxide and metallic states when calculating formation enthalpy.

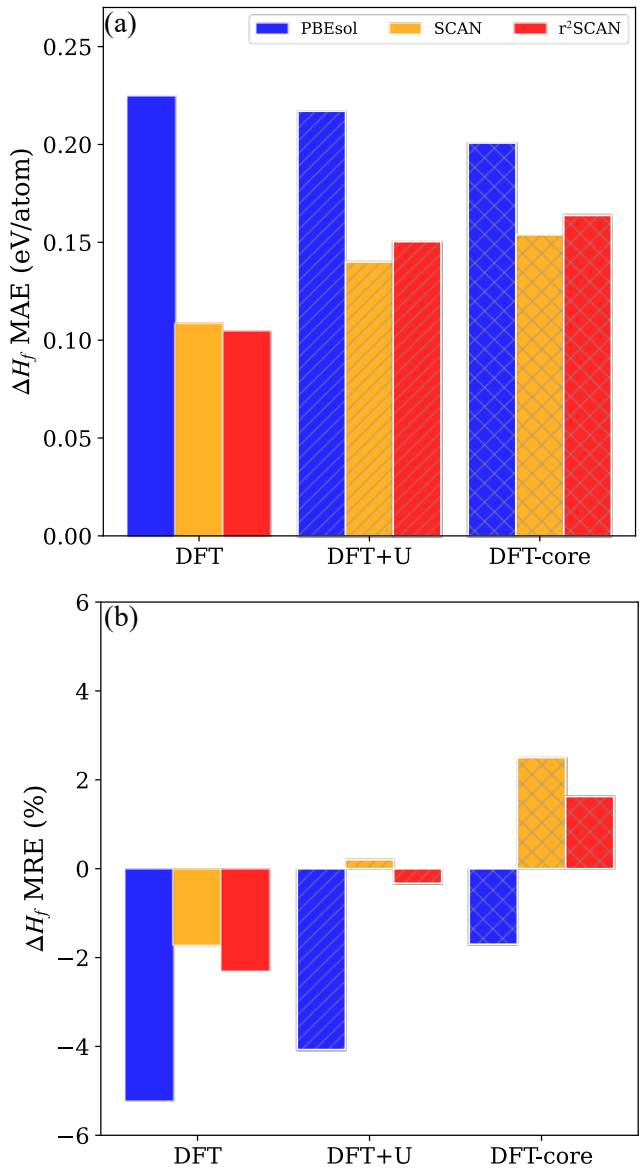

FIG 7. (a) The MAE and (b) the MRE of the DFT, DFT+$U$, and DFT-core formation enthalpies $\Delta H_f$ compared to experiment for 15 REOs using PBEsol, SCAN, and r²SCAN. Note that DFT-core does not include predictions for $Y_2O_3$, $La_2O_3$, $PrO_2$, $CeO_2$.

Figure 7 quantifies the errors in formation energies for DFT, DFT+$U$, and DFT-core in comparison to experimental values [123,124]. PBEsol consistently underestimates experimental values, except for PrO$_2$, resulting in a negative MRE of 5.22% [Fig. 7(b)] and an MAE of 225 meV/atom [Fig. 7(a)]. The systematic underestimation of formation energies for PBEsol is a result of the well-known GGA overbinding error in the O$_2$ molecule that corresponds to an underbinding of the compound with respect to the elements [60]. SCAN and r$^2$SCAN without +$U$ perform similarly, with MAEs (MREs) of 109 meV/atom (-1.72%) and 104 meV/atom (-2.31%), respectively. Although they present slight underestimation [Fig. 7(b)], meta-GGA functionals are a significant improvement over PBEsol for formation enthalpy predictions.

The addition of Hubbard $U$ slightly improves PBEsol+$U$ errors to 218 meV/atom but worsens SCAN+$U$ and r$^2$SCAN+$U$ MAEs. The meta-GGA+$U$ MRE decrease reflects the fortuitous shift in nearly half the meta-GGA's $\Delta H_f$ to above the line of parity [Fig. S7]. The $f$-core pseudopotentials yield more accurate $\Delta H_f$ energies with lower MAEs than PBEsol and PBEsol+$U$ but do not demonstrate improvements over meta-GGA or meta-GGA+$U$. DFT-core, despite frequent use based on the assumption of limited $f$-electron involvement in bonding, does not exhibit improved MAEs compared to $\Delta H_f$ energies calculated from $f$-band meta-GGAs. SCAN and r$^2$SCAN, without +$U$, outperform $f$-core with around 30% and 35% lower MAEs, respectively.

Table IV contains the relative stability between the B-type and C-type polymorphs $\Delta E_{\text{B-type–C-type}}$ of Sm$_2$O$_3$, Eu$_2$O$_3$, and Gd$_2$O$_3$. When assessing the relative stabilities of Sm oxides, DFT favors C-Sm$_2$O$_3$, while DFT+$U$ predicts the B-type polymorph to be more stable. The discrepancies in predicting ground state polymorphs arise from differences in the qualitative description of electronic structure, leading to errors in evaluated energies [61]. DFT properly predicts the C-type polymorph as more stable because both structural calculations are equally influenced by SIEs in the absence of +$U$. However, in the presence of +$U$, the C-type structure is dominated by relativistic effects, necessitating the use of SOC for an accurate representation of the band gap characteristics. This is supported by the observation that the $\Delta E_{\text{B-type–C-type}}$ from both SCAN+$U$ and the 52 version of PAW PBEsol+$U$ predict C-Sm$_2$O$_3$ as the ground state. The key distinction is that these approaches achieve a qualitatively accurate density of states without relying on SOC. When the qualitative description of electronic structure is correct, proper predictions of the ground state polymorphs are obtained.

TABLE IV. Relative stability $\Delta E_{\text{B-type–C-type}} = E_{\text{B-type}} - E_{\text{C-type}}$ (in meV/atom) between the B-type and C-type polymorphs of Sm, Eu, Gd. Positive $\Delta$ indicates C-type is more stable.

| Compound | DFT | | | DFT+$U$ | | |
|---|---|---|---|---|---|---|
| | PBEsol | SCAN | r$^2$SCAN | PBEsol | SCAN | r$^2$SCAN |
| Sm$_2$O$_3$ | 27.3 | 24.4 | 12.1 | -12.5 (86.5[a]) | 3.19 | -61.6 |
| Eu$_2$O$_3$ | 26.9 | 26.7 | 28.9 | 22.8 | 24.2 | 26.9 |
| Gd$_2$O$_3$ | 20.7 | 19.7 | 22.0 | 21.7 | 19.2 | 22.2 |

[a]Using 54 PAW.

## D. Computational efficiency

We quantify the computational time for PBEsol, SCAN, and r$^2$SCAN, employing both $f$-band and $f$-core approaches with either the 54 or 64 versions of PAW. Our objective is to provide a qualitative understanding of the relative computational costs across the considered frameworks, rather than offer precise measurements of computational resources for each XC framework. The data presented in Fig. 8 is generated through separate calculations, utilizing converged structures, FM ordering, and excluding the application of Hubbard $U$. For consistency, we performed a single ionic step (NSW=0) with uniform parallelization settings, maintaining the same number of nodes (1), cores (20), and multiprocessing task settings.

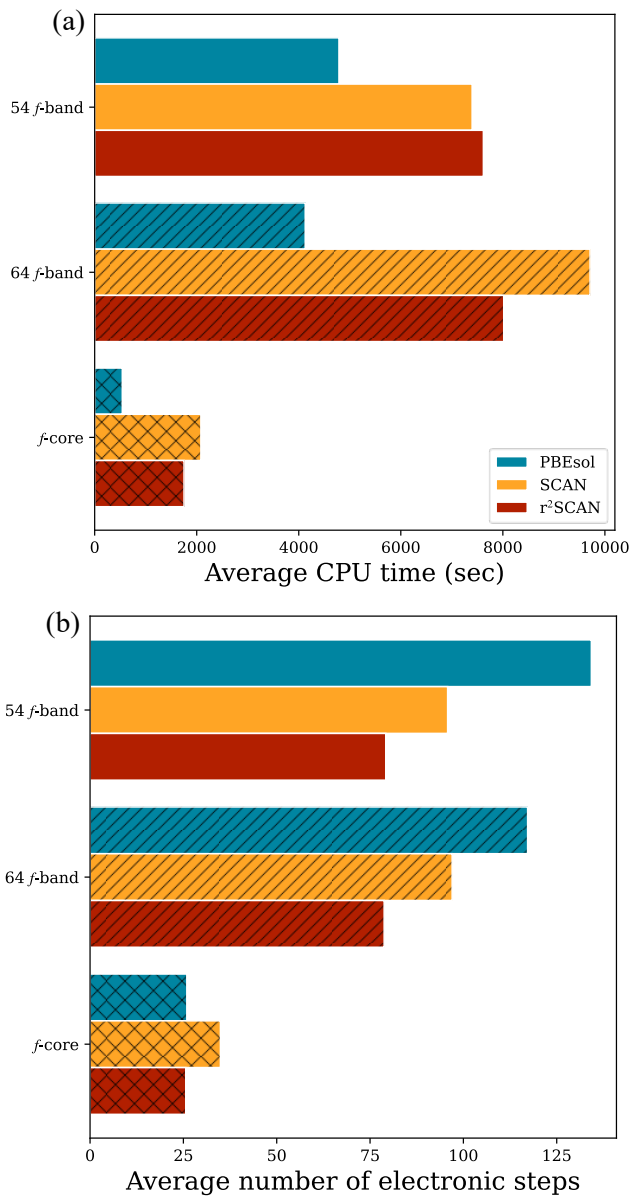

FIG 8. (a) Average computational time and (b) average number of electronic steps of PBEsol (blue), SCAN (orange), and r$^2$SCAN (red) using the $f$-band and $f$-core pseudopotentials. All data used for comparison was without Hubbard $U$ corrections in ferromagnetic ordering.

Figure 8 illustrates the total average CPU time [Fig. 8(a)] and the average number of electronic steps, i.e., self-consistent field (SCF) cycles [Fig. 8(b)]. Both meta-GGA functionals require approximately 2.0× the CPU time than PBEsol when using the 54 and 64 PAW versions. As expected, the $f$-core approach demonstrates faster convergence with fewer electronic steps than the $f$-band method. Compared to the $f$-core approach, the $f$-band method with 64 PAW version takes 7.6× (PBEsol), 4.7× (SCAN), and 4.6× (r$^2$SCAN) more CPU time to converge the single ionic step, and utilizes 4.3× (PBEsol), 2.5× (SCAN), and 3.1× (r$^2$SCAN) more SCF cycles. Among the $f$-band approaches, r$^2$SCAN expends the least SCF cycles with $f$-core necessitating only a fraction of SCF cycles compared to $f$-band. While our evaluation focuses on a specific subset of materials, the observed relative computational efficiency trends for REOs align with those derived from a broader and more diverse dataset [62]. In assessing the computational requirements of AFM ordering, we find that broken-symmetry magnetic ordering consumes more CPU time and requires a greater number of SCF cycles [Fig. S17].

**IV. DISCUSSION**

Our comprehensive evaluation delves into the performance of thirteen different exchange-correlation approximations on rare earth oxides, assessing their efficacy across various aspects of structural, magnetic, and electronic properties. It is crucial to understand the limitations of each method as their predictive capabilities vary across element types and investigated properties.

We find that the meta-GGA functionals show a clear improvement of binary REO lattice volumes in comparison to PBEsol with r$^2$SCAN predicting the largest volumes and having the lowest MAE [Fig. 2]. While this finding is similar to that observed by both Kothakonda *et. al.* [64] and Isaacs and Wolverton [60], it differs from the results of Kingsbury *et. al.*'s [62] that PBEsol performs best for 'strongly bound' materials, i.e., materials with formation energies lower than -1 eV/atom. This difference may be attributed to our focus on REOs as opposed to materials spanning across the periodic table. Ultimately, the meta-GGA functionals' ability for improved structural predictions can be attributed to the reduced self-interaction errors. Introducing Hubbard $U$ further reduces delocalization errors by facilitating enhanced electron localization on the RE $d$/$f$ orbitals resulting in consistent increases in lattice volume. As SCAN and PBEsol underpredict lattice volumes, the consistent increase in DFT+$U$ structural properties incidentally reduces PBEsol+$U$'s and SCAN+$U$'s errors while increasing the errors of r$^2$SCAN+$U$ as it already has accurate structural predictions without any corrections. In the context of DFT+$U$ approaches, the specific properties' enhancement or deterioration is dependent on the Hubbard $U$ value. Employing a single $U$ value across different properties and polymorphs has consequences, such as the non-transferability of Hubbard $U$, where improvement in one property often comes at the expense of another. Our focus was not on determining an optimal $U$ value, but rather on elucidating the trends and influences of +$U$ on the predicted properties of meta-GGA functionals.

The lattice volumes predicted using the $f$-core potentials did not show improvement over the $f$-band potentials. DFT-core had higher volume MAEs than all $f$-band approaches, DFT or DFT+$U$, except for PBEsol-core which had a better MAE than PBEsol. Yet once delocalization errors are corrected with +$U$, PBEsol+$U$ outperforms DFT-core. The lower errors of $f$-band predictions compared to $f$-core indicate the non-negligible influence of the localized $f$-electrons on bonding, in contrast to the common misconception that the inner $f$-electrons do not contribute to bonding in REOs [125]. The 4$f$ contribution to bonding can be expected because 4$f$ electrons are similar in energy to the O-2$p$ valence electrons. The presence of appropriately localized valence $f$-electrons leads to the best structural predictions. While r$^2$SCAN would be the optimal choice for structural predictions of REOs, PBEsol+$U$ can be considered a compromise of accuracy and cost as it has only approximately

12% higher atomic volume MAE than SCAN but requires less than half its computational cost. Although r$^2$SCAN-core has an MAE more than 80% higher than r$^2$SCAN, it can maintain the periodic lanthanide contraction with only a 25% larger MAE than PBEsol+$U$ for a fraction of the computational cost.

Focusing on electronic structure, GGA methods like PBEsol are typically known to have unphysical descriptions of the 4$f$ states because the partially occupied 4$f$ orbitals get pinned at the Fermi level [126]. We likewise find that the meta-GGAs can still be at the mercy of improper descriptions of 4$f$ states. When relativistic effects have low influence, the inclusion of the Hubbard $U$ parameter can remedy the flawed description of band gaps. We note that careful consideration of the chosen $U$ value is crucial to avoid over-localization, affecting the band gap structure. For instance, in A-Nd$_2$O$_3$, an excessively large $U$ value can push the $f^n$ state into the valence band, resulting in a band gap structure akin to Fig. 3(a) rather than Fig. 3(b). This proximity of the filled $f^n$ band to the valence band elucidates why Nd necessitates a lower $U$ value compared to its RE neighbors. These considerations underscore the importance of thoughtfully selecting $U$ values to accurately capture the electronic structure of REOs and prevent unintended distortions in band gap characteristics. Further, we find that crystal symmetry can impact the level of theory necessary for electronic predictions. This is emphasized by the predictions between the polymorphs of Sm$_2$O$_3$ and Ce$_2$O$_3$. At the DFT level, the meta-GGA functionals produced insulating character for A-Ce$_2$O$_3$, but DFT+$U$ was needed for C-Ce$_2$O$_3$. Similarly, B-Sm$_2$O$_3$ did not necessitate SOC for proper band gap character [Fig. S12]. The difference in crystal symmetry between the two polymorphs of Sm$_2$O$_3$ may contribute to a stronger presence of relativistic effects in the C-type. This is distinct from Eu$_2$O$_3$ where the DFT+$U$ level is sufficient for proper DOS for both polymorphs.

Residing one rung higher than PBEsol on the XC ladder, SCAN and r$^2$SCAN have shown slightly more realistic band gaps than those of GGAs [64,67], but neither SCAN nor r$^2$SCAN, consistently predict band edge orbital character across all REOs, in agreement with Ref. [33]. The meta-GGA functionals provide a significant advantage over PBEsol for the electronic properties of A-Ce$_2$O$_3$, A-Nd$_2$O$_3$, and B-Sm$_2$O$_3$ in which both meta-GGA functionals properly predict electronic behavior and band gap character without the need for empirical corrections. Apart from those instances, PBEsol and its +$U$ and SOC derivatives performs as well as the meta-GGA functionals. While producing insulating behavior, the $f$-core model fails to qualitatively reproduce band gap characteristics and would not be ideal for predicting electronic behavior.

Moreover, the meta-GGA functionals do not preclude the need to address relativistic effects through SOC when electronic properties are dominated by their influence. One can determine if a calculation needs SOC by whether the DFT energy is lowered when compared to without SOC, i.e. a collinear magnetic compound should have the same DFT energy with SOC [85]. Including SOC to a calculation can lead to significant changes to predicted properties such as larger band gaps, more favorable DFT energies, and enhanced magnetic moments. Although we do observe these changes for other 4$f$ compounds to a lesser degree, like Nd$_2$O$_3$ and Eu$_2$O$_3$, qualitatively appropriate electronic structures are achieved without using SOC in contrast to PrO$_2$, A-Pr$_2$O$_3$, and C-Sm$_2$O$_3$. Even though a non-zero band gap may be predicted, the band edge orbital characters can be wrongly assigned due to invalid localization of 4$f$ bands near the Fermi level. This suggests that Hund's rule and crystal field splitting are not well represented by any of the functionals for the cases of PrO$_2$, A-Pr$_2$O$_3$, and C-Sm$_2$O$_3$ [33]. Thus, SIEs of the REOs with their partially filled 4$f$ states are not properly amended by the meta-GGA functionals, and consequently, any analysis of predicted electronic character, especially without SOC for compounds containing Pr and Sm, should carefully consider these limitations.

Similar to previous reports, we observe both SCAN and r$^2$SCAN have much more accurate formation energies than PBEsol because they are unencumbered by the well-known O$_2$ over-binding error. Among our considered approximations, r$^2$SCAN on average produces the most accurate REO $\Delta H_f$ energies in comparison to experiment. The minor improvement (PBEsol) and worsening (meta-

GGAs) of the average $\Delta H_f$ MAE with +$U$ may be attributable to our use of an average $U$ value, and employing a specifically optimized $U$ for calculating $\Delta H_f$ could rectify this reduced agreement with experiment [60,63,65]. For all functionals, the largest deviations in $\Delta H_f$ predictions versus experiment are observed for $PrO_2$, $Sm_2O_3$, and $Eu_2O_3$. In comparing PAW version 54 with 64, there are significant decreases in the MAE for $PrO_2$ and $Eu_2O_3$ listed in Table S5 and Table S4, respectively. This indicates that the large errors could be due, in part, to the descriptions within the pseudopotentials utilized. We also observe improvements of MAE by as much as 5% when including SOC, insinuating the level of theory influencing accuracy.

Freezing the $f$-electrons in the core ($f$-core) comes with great computational speed-up and essentially removal of large delocalization errors as the $f$-electrons are no longer considered in the valence band. Computational expenses can be mitigated by employing SCAN-core and r$^2$SCAN-core as they use barely a quarter of the time compared to their $f$-band counterparts. This speedup comes with an over 25% increase in $\Delta H_f$ MAEs, yet their MAEs are still superior to the slower PBEsol+$U$. Ultimately, r$^2$SCAN is the optimal choice for predictions of $\Delta H_f$, keeping in mind that proper qualitative descriptions of electronic structure may be necessary to get correct predictions of relative stabilities for ground state polymorphs.

## V. Conclusions

Conventional DFT approximations struggle to accurately describe strong electron correlation and localization in 4$f$ orbitals, which is critical to predicting structure-property relationships in REOs and to advancing their development in various applications. We have benchmarked fundamental properties of light REOs across the PBEsol, SCAN, and r$^2$SCAN exchange-correlation functionals, while also examining the effects of Hubbard $U$ and SOC corrections. Our findings provide practical guidance on selecting DFT functionals for studying REOs, considering the trade-off between computational cost and accuracy.

The meta-GGA functionals, with their reduced self-interaction and $O_2$ binding errors, exhibit improved volume and formation enthalpy predictions. For structural predictions, r$^2$SCAN is the most accurate choice, but PBEsol+$U$ offers a favorable balance between accuracy and computational efficiency. Similarly, r$^2$SCAN is ideal for energetic predictions, although r$^2$SCAN-core can be utilized when minimizing computational expense. Our bandgap analysis reveals that the meta-GGA functionals generally perform meaningfully better than PBEsol only in specific cases, such as A-$Ce_2O_3$, A-$Nd_2O_3$, and B-$Sm_2O_3$. For most other systems, accurate band gap predictions require either +$U$ or +$U$+SOC, underscoring the persistence of delocalization errors and lack of relativistic effects. Therefore, PBEsol combined with either +$U$ or +$U$+SOC is an amendable compromise for electronic properties.

In sum, r$^2$SCAN is recommended overall for the highest accuracy in structural, electronic, and energetic predictions at only about 2.0× more the CPU time of PBEsol. When computational speed is a priority, PBEsol+$U$ could be used for structural predictions, and r$^2$SCAN-core for energetic predictions. For electronic properties, PBEsol combined with +$U$ or +$U$+SOC is a sensible selection for a more balanced approach between speed and accuracy, with the understanding that DFT has inherent limitations in capturing the complex electronic structure of REOs.

**Acknowledgements**
The authors acknowledge the use of facilities and instrumentation supported by NSF through the Pennsylvania State University Materials Research Science and Engineering Center [DMR-2011839] Computations for this research were performed on the Pennsylvania State University's Institute for Computational and Data Sciences' Roar supercomputer.

# SUPPLEMENTAL MATERIAL

# Performance of Exchange-Correlation Approximations to Density-Functional Theory for Rare-earth Oxides

[*,1]Mary Kathleen Caucci, [1]Jacob T. Sivak, [2]Saeed S.I. Almishal, [3]Christina M. Rost, [2]Ismaila Dabo, [2] Jon-Paul Maria, [*,1,2,4,5]Susan B. Sinnott

[1]*Department of Chemistry, The Pennsylvania State University, University Park, PA 16802*
[2]*Department of Materials Science and Engineering, The Pennsylvania State University, University Park, PA 16802*
[3]*Department of Materials Science and Engineering, Virginia Polytechnic Institute and State University, Blacksburg, VA 24060*
[4]*Materials Research Institute, The Pennsylvania State University, University Park, PA, 16802*
[5]*Institute for Computational and Data Science, The Pennsylvania State University, University Park, PA, 16802*

(Dated: August 19, 2024)

**Section S1. Rare-earth metals**

*A. Lattice volumes*

Figure S1 displays the lattice volume predictions of RE metals for DFT, DFT+U, and DFT-core approaches, and Fig. S2 shows the MAE and MRE compared to experiment [1,2]. The approximation with the lowest MAE for RE metallic lattice volumes was SCAN 2.84 Å$^3$/atom, followed closely by r$^2$SCAN and r$^2$SCAN-core which both have an MAE of 2.94 Å$^3$/atom. Similar to the REOs, the MAE for DFT+$U$ improves for PBEsol+U, but increases for r$^2$SCAN+$U$. When comparing the MAEs of $f$-core to their relevant $f$-band predictions, the meta-GGA $f$-core predictions are better than their $f$-band, while both PBEsol+$U$ and PBEsol-core showed the same extent of improvement over PBEsol.

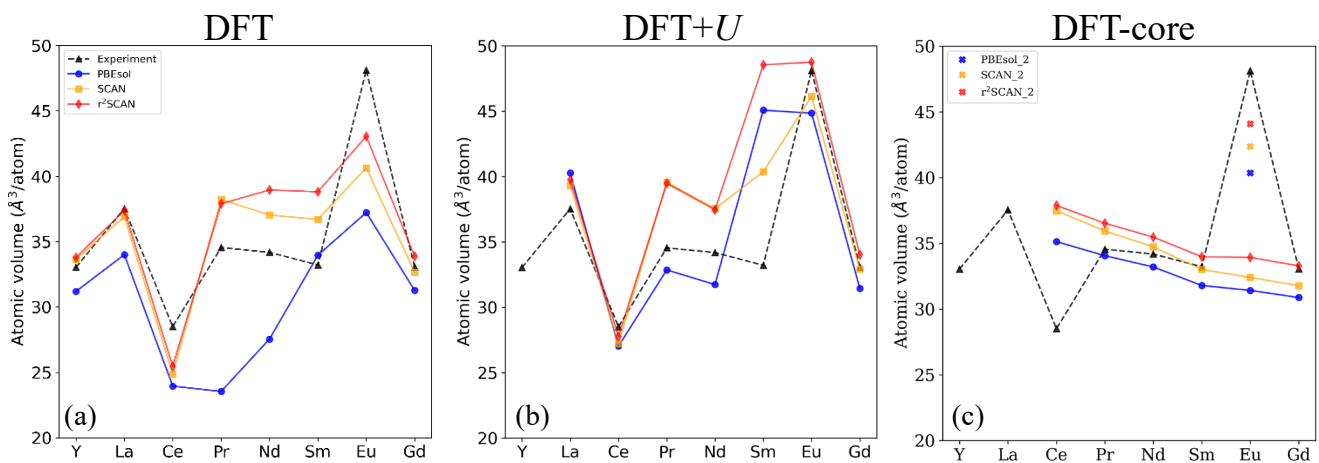

FIG S1. Predicted lattice volumes for RE metals using (a) DFT, (b) DFT+$U$, (c) DFT-core methods. In (c) the cross points represent values calculated with 'Eu_2' pseudopotential rather than 'Eu_3'.

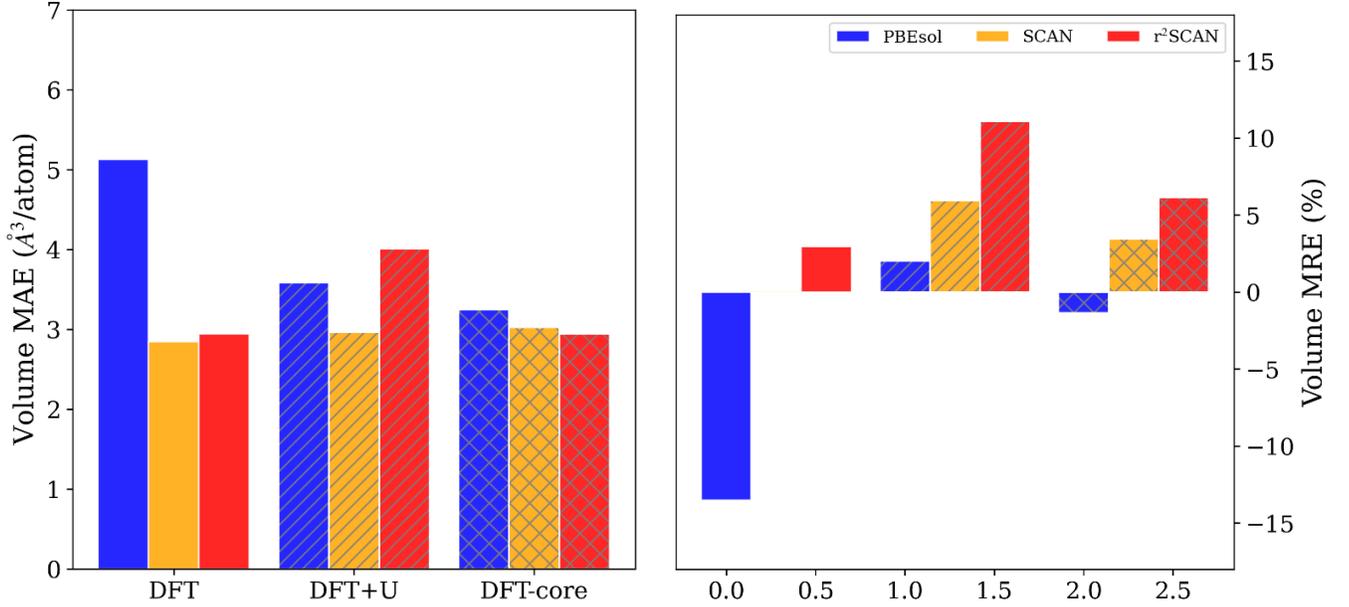

FIG. S2. The mean absolute error, MAE, (left) and mean relative error, MRE, (right) for DFT, DFT+$U$, and DFT-core lattice volume predictions compared to experiment for RE metals. Note that $f$-core pseudopotentials do not include predictions for Y and La.

Looking at Fig. S1, we can assess the various approximations' ability to accurately capture the metallic RE periodic trend. Since we take the α-Ce rather than γ-Ce as the ground state, there is a drop in the atomic volume, which all the approximations were able to describe except for DFT-core. The substantial increase of atomic volume for Eu is due to its preference for forming the divalent cubic structure [3]. This is evident in Fig. S1(c) by the Eu_2 pseudopotential having considerable improvement over the analogous Eu_3 pseudopotential, however the Eu_2 potentials do not have any substantial improvement over the respective DFT $f$-band model in capturing this increase in atomic volume.

## B. Electronic Structure

When comparing the predicted effective magnetic moments of metallic rare earth elements in comparison to experiment [Fig. S2], there are notable deviations from experiments, particularly for Nd, Sm, and Eu, which have been previously noted [3–5]. These deviations are unsurprising when considering the complex magnetic states of these three elements cannot be easily captured using standard DFT approaches, in addition to challenges in experimental investigations of pure lanthanides [3,6–9]. SCAN consistently predicted the largest ASMs both with and without a +U correction, other than Eu with r$^2$SCAN. For all the functionals, when going from DFT to DFT+$U$, the ASMs increased for Gd but decreased for Pr and Eu. While for Nd (Sm), the ASMs decreased (increased) for PBEsol and r2SCAN but increased (decreased) for SCAN. On average, there is minimal quantitative influence of +$U$ correction on the effective magnetic moments, where only a few of the relative differences between DFT and DFT+$U$ were larger than 1%. Those higher relative differences were for Nd (-3.79%) and Sm (2.92%) from r$^2$SCAN as well as Nd (-1.41%) and Gd (2.38%) from PBEsol. Incorporating SOC shows slight improvement for Pr and Nd, but not for Sm and Eu where including orbital polarization could help [4]. The net ionic charges of the metallic state are zero as there is no transfer of electrons.

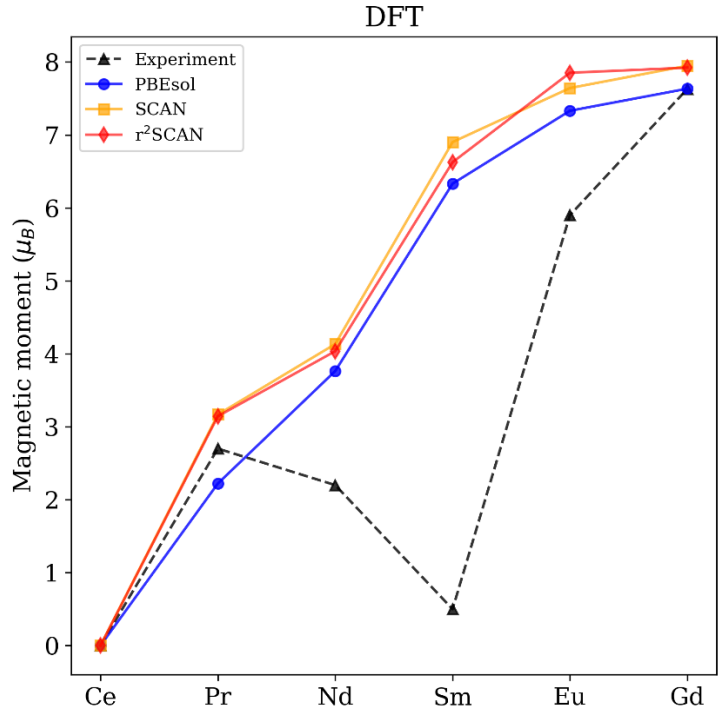

FIG. S3. The calculated average total magnetic moment (in $\mu_B$) for the RE metals compared to experiment using DFT with either PBEsol (blue circles), SCAN (yellow squares), or r$^2$SCAN (red diamonds) XC functionals.

## Section S2. Hubbard U transferability

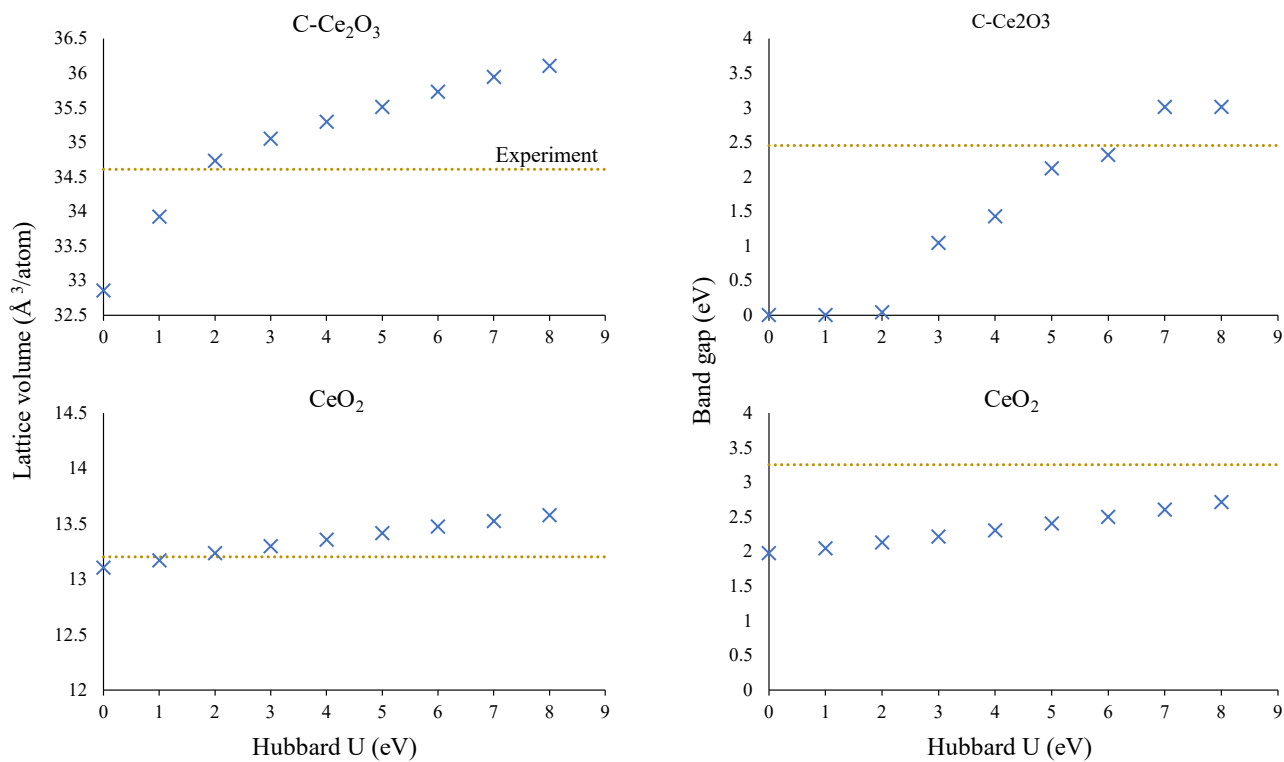

FIG S4. Calculated lattice parameters (right) in Å and band gaps (left) in eV for the fluorite (bottom) and bixbyite (top) structures at increasing U values (in eV) with experimental (orange dotted line). It can be noted that at higher U values, band gaps are in better agreement with experiment, but better electronic accuracy comes at the cost of structural accuracy with an overestimated lattice parameter.

## Section S3. Volumes

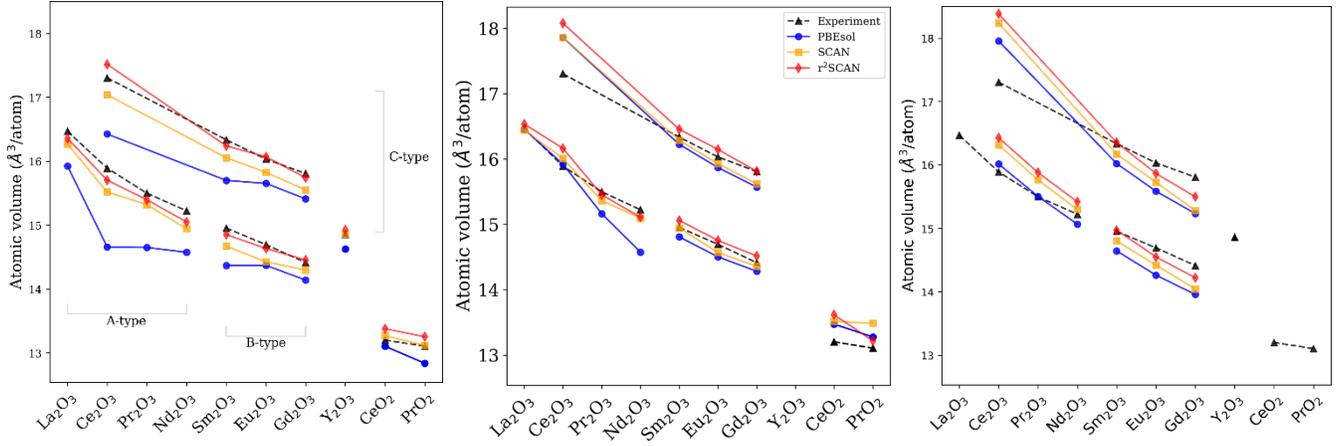

FIG S5. Predicted lattice volumes of REOs compared to experiment for (a) DFT, (b) DFT+U, (c) DFT-core methods.

TABLE S1. Relative errors of lattice volumes compared to experimental values for selected compounds. Negative relative errors indicate underpredicted values in relation to experiment.

| Compound | Theory | DFT | DFT+SOC | DFT+$U$ | DFT+$U$+SOC | $f$ in core |
|---|---|---|---|---|---|---|
| B-$Eu_2O_3$ | PBEsol | -2.20% | -2.23% | -1.30% | -1.26% | -2.93% |
|  | SCAN | -1.82% |  | -0.84% |  | -1.86% |
|  | $r^2$SCAN | -0.40% | -0.39% | 0.41% | 0.02% | -0.97% |
| C-$Sm_2O_3$ | PBEsol | -3.89% | -3.77% | -0.68% | -0.90% | -1.92% |
|  | SCAN | -1.81% |  | -0.29% |  | -1.01% |
|  | $r^2$SCAN | -0.59% | -0.68% | 0.73% | 0.70% | 0.11% |
| A-$Pr_2O_3$ | PBEsol | -5.49% | -6.52% | -2.17% | -2.17% | 0.03% |
|  | SCAN | -1.59% |  | -0.88% |  | 1.71% |
|  | $r^2$SCAN | -0.66% | -1.11% | -0.31% | 0.12% | 2.45% |
| $PrO_2$ | PBEsol | -2.04% | -2.25% | 1.31% | -0.78% | 14.29% |
|  | SCAN | 0.09% |  | 2.99% |  | 16.63% |
|  | $r^2$SCAN | 1.13% | 1.42% | 0.85% | -1.88% | 17.70% |
| $CeO_2$ | PBEsol | -0.74% |  | 2.05% |  | 17.17% |
|  | SCAN | 0.53% |  | 2.37% |  | 19.47% |
|  | $r^2$SCAN | 1.34% |  | 3.13% |  | 20.41% |

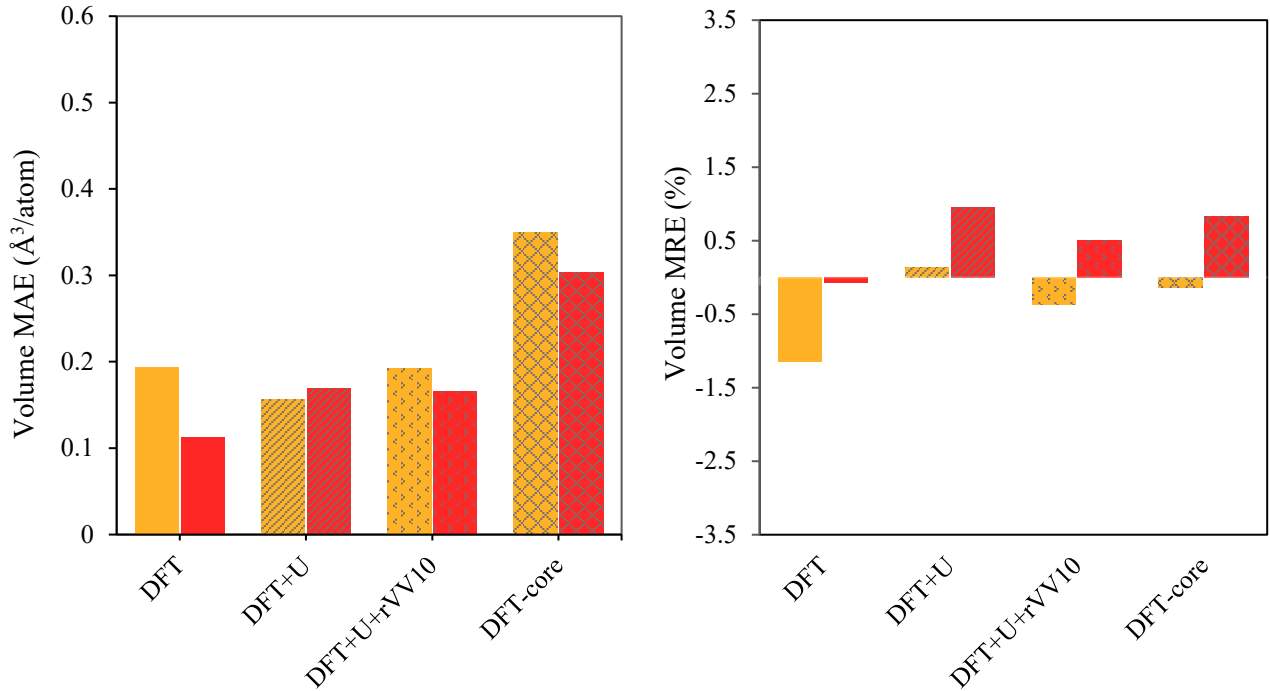

FIG. S6. The mean absolute error, MAE, (left) and mean relative error, MRE (right) for DFT, DFT+$U$, DFT+$U$+rVV10, and DFT-core lattice volume predictions compared to experiment using either SCAN (yellow) or r$^2$SCAN (red).

**Section S4. Formation Enthalpy**

There will inevitably be systematic errors due to thermal expansion when comparing experimental room temperature volumes with T= 0K theoretical values. As emphasized in work by Delin et. al., it's expected that these effects are very small [10,11]. The REOs in our study have an average coefficient of 10.2 x10$^{-6}$ α °C$^{-1}$ [12] which is a volume thermal expansion of about 1%. Therefore, it's reasonable to not consider these effects in our formation enthalpies.

TABLE S2. Thermal expansion coefficients for REOs of interest.

| Compound | Thermal expansion coefficient per °C (x10$^{-6}$) |
|---|---|
| $Y_2O_3$ | 8.0 |
| $CeO_2$ | 11.6 |
| $Pr_6O_{11}$ | 15.2 |
| $Nd_2O_3$ | 12.7 |
| $Sm_2O_3$ | 8.5 |
| $Eu_2O_3$ | 7.2 |
| $Gd_2O_3$ | 8.1 |

TABLE S3. Solid elemental reference states [1].

| Symbol | Spacegroup | Structure |
|--------|------------|-----------|
| Y  | $P6_3/mmc$ | hcp |
| La | $P6_3/mmc$ | dhcp |
| Ce | $Fm\bar{3}m$ | fcc |
| Pr | $P6_3/mmc$ | dhcp |
| Nd | $P6_3/mmc$ | dhcp |
| Sm | $R\bar{3}m$ | rhomb, αSm-type |
| Eu | $Im\bar{3}m$ | bcc |
| Gd | $P6_3/mmc$ | hcp |

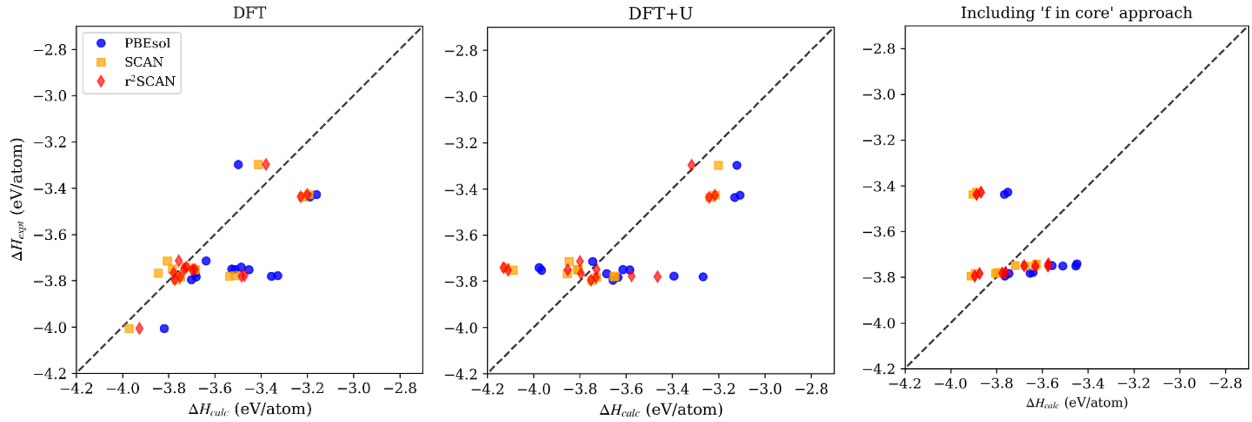

FIG S7. Comparison of calculated and experimental formation energy $\Delta H_f$ for 15 REOs using PBEsol (blue circle), SCAN (yellow square), and r$^2$SCAN (red diamond). Black dashed line in (a) and (b) is parity between calculated and experimental values. C-$Y_2O_3$ is not included in DFT+U.

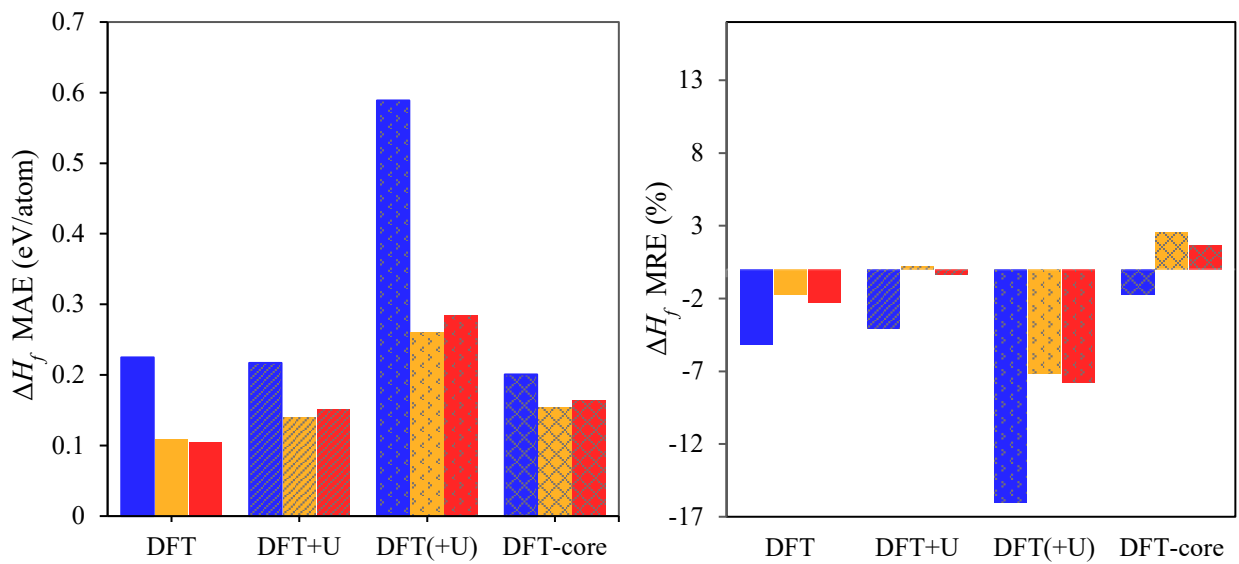

Fig. S8. The MAE (left) and the MRE (right) of the DFT, DFT+$U$, DFT(+$U$) and DFT-core formation enthalpies $\Delta H_f$ compared to experiment for 15 REOs using PBEsol (blue), SCAN (yellow), and r$^2$SCAN (red). DFT(+$U$) indicates $\Delta H_f$ calculated without $U$ on the metallic reference state. The increase in mean errors associated with mixing DFT and DFT+U energies may be improved by determining an energy correction term. Note that DFT-core does not include predictions for Y$_2$O$_3$, La$_2$O$_3$, PrO$_2$, CeO$_2$.

TABLE S4. Relative errors of formation enthalpy $\Delta H_f$ compared to experimental values for selected compounds.

| Compound | Theory | DFT | DFT+SOC | DFT+$U$ | DFT+$U$+SOC | $f$ in core |
|---|---|---|---|---|---|---|
| B-$Eu_2O_3$ | PBEsol | -7.79% | -7.34% | -9.34% | -8.31% | 9.46% |
| | SCAN | -6.85% | | -6.18% | | 13.43% |
| | r$^2$SCAN | -6.65% | -6.59% | -6.17% | -4.30% | 12.86% |
| C-$Sm_2O_3$ | PBEsol | -11.25% | - | -13.56% | -8.89% | -3.38% |
| | SCAN | -6.44% | | -3.40% | | 0.65% |
| | r$^2$SCAN | -7.86% | -7.10% | -8.39% | -4.54% | -0.14% |
| A-$Pr_2O_3$ | PBEsol | -6.35% | -5.94% | -4.46% | -3.70% | 6.35% |
| | SCAN | -1.77% | | 1.45% | | 2.08% |
| | r$^2$SCAN | -1.71% | -1.58% | 2.70% | -0.38% | 3.21% |
| $PrO_2$ | PBEsol | 6.12% | 5.53% | -5.36% | -1.77% | -15.67% |
| | SCAN | 3.46% | | -2.94% | | -12.89% |
| | r$^2$SCAN | 2.46% | 2.79% | -0.56% | -2.38% | -14.32% |
| $CeO_2$ | PBEsol | -2.16% | | -2.20% | | -27.36% |
| | SCAN | 2.06% | | 2.34% | | -24.65% |
| | r$^2$SCAN | 0.32% | | 0.73% | | -26.12% |

## A. Pseudopotentials

Part of the 64 PBE PAW potential set included new lanthanide potentials with an "_h" suffix. These potentials use a smaller core radius of 2.2 a.u. than the previous versions, as well as placing 0.5 electrons from the $4f$ shell into the $5d$ shell. More information regarding these potentials can be found in Ref. [13] and the VASP wiki page. Future efforts towards validation of these hard pseudopotentials recommended by Ref. [13] are necessary.

TABLE S5. Relative errors of formation enthalpy $\Delta H_f$ compared to experimental values for selected compounds using 54 PAW.

| Compound | Theory | DFT | DFT+$U$ |
|---|---|---|---|
| B-$Eu_2O_3$ | PBEsol | -10.55% | -18.02% |
|  | SCAN | -10.97% | -13.50% |
|  | r$^2$SCAN | -10.69% | -12.93% |
| C-$Sm_2O_3$ | PBEsol | -10.30% | -13.26% |
|  | SCAN | -5.40% | -6.48% |
|  | r$^2$SCAN | -7.37% | -8.74% |
| A-$Pr_2O_3$ | PBEsol | -6.07% | -8.01% |
|  | SCAN | 0.25% | -1.93% |
|  | r$^2$SCAN | -1.67% | -8.45% |
| $PrO_2$ | PBEsol | 7.39% | -4.54% |
|  | SCAN | 5.84% | -4.63% |
|  | r$^2$SCAN | 5.07% | -3.14% |

# Section S5. Band gaps

TABLE S6. Relative errors of band gaps compared to experimental values for selected compounds.

| Compound | Theory | DFT | DFT+SOC | DFT+U | DFT+U+SOC | $f$ in core |
|---|---|---|---|---|---|---|
| B-Eu$_2$O$_3$ | PBEsol | -99.99% | -100.00% | -45.92% | -44.06% | -13.37% |
|  | SCAN | -99.98% |  | -46.03% |  | -0.80% |
|  | r$^2$SCAN | -99.98% | -99.98% | -48.05% | - | -0.37% |
| C-Sm$_2$O$_3$ | PBEsol | -99.98% | -99.98% | -57.18% | -35.64% | -24.71% |
|  | SCAN | -88.80% |  | -35.10% |  | -15.37% |
|  | r$^2$SCAN | -91.50% | -81.81% | -55.55% | -30.73% | -14.34% |
| A-Pr$_2$O$_3$ | PBEsol | -98.91% | -99.99% | -35.70% | -38.73% | -5.70% |
|  | SCAN | -85.50% |  | -29.53% |  | 5.48% |
|  | r$^2$SCAN | -84.35% | -75.25% | -32.56% | -27.74% | 5.73% |
| PrO$_2$ | PBEsol | -99.88% | -99.93% | -99.05% | -21.41% | -99.57% |
|  | SCAN | -99.98% |  | -99.64% |  | -99.78% |
|  | r$^2$SCAN | -99.88% | -55.91% | -45.00% | -28.10% | -99.80% |
| CeO$_2$ | PBEsol | -39.16% |  | -23.14% |  | -99.85% |
|  | SCAN | -31.02% |  | -19.53% |  | -99.85% |
|  | r$^2$SCAN | -33.11% |  | -16.60% |  | -95.74% |

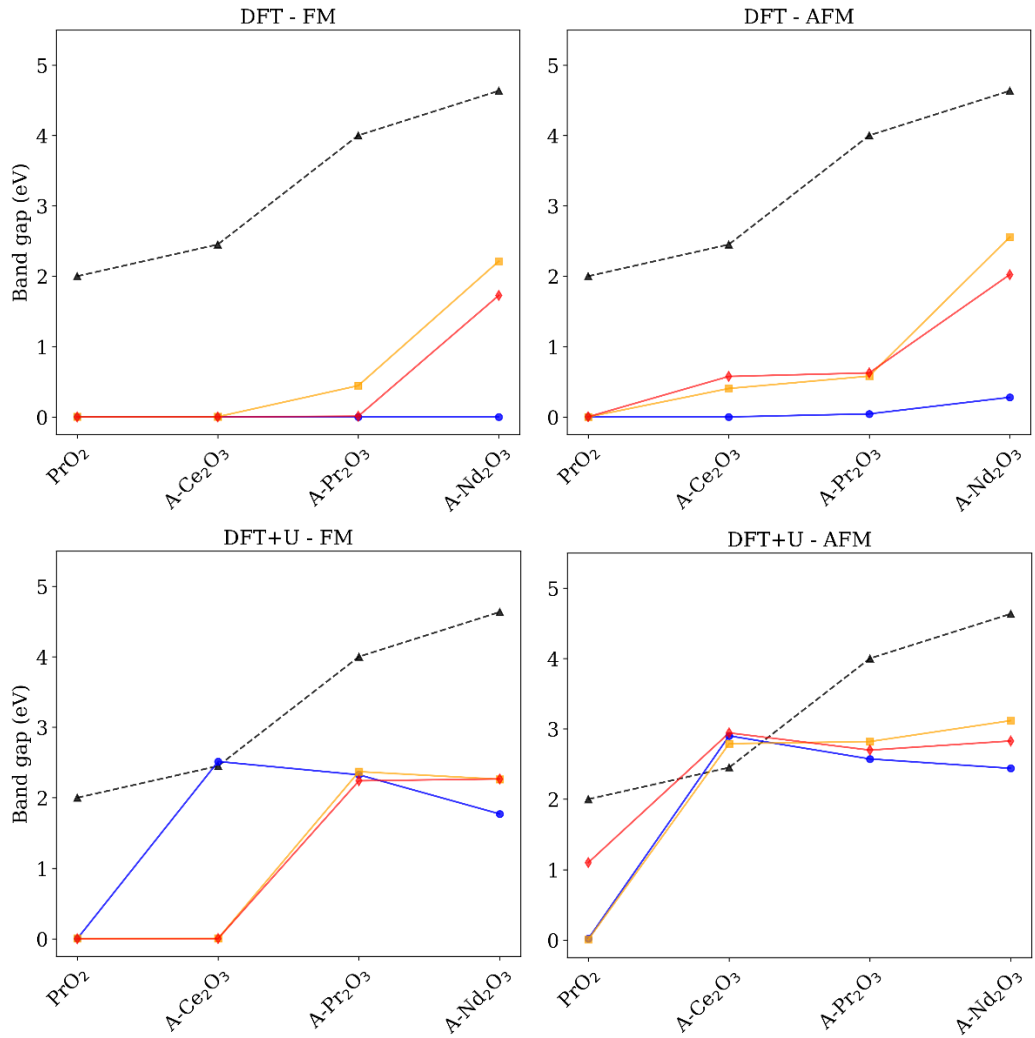

Fig. S9. Comparison of experimental band gaps to predicted minimum electronic band gaps as obtained using PBEsol (blue circles), SCAN (yellow squares), and r²SCAN (red diamonds) in (a) DFT – FM, (b) DFT – AFM, (c) DFT+U – FM, and (d) DFT+U – FM. We find that there are larger band gaps when in the AFM ordering.

## Section S6. Partial density of states

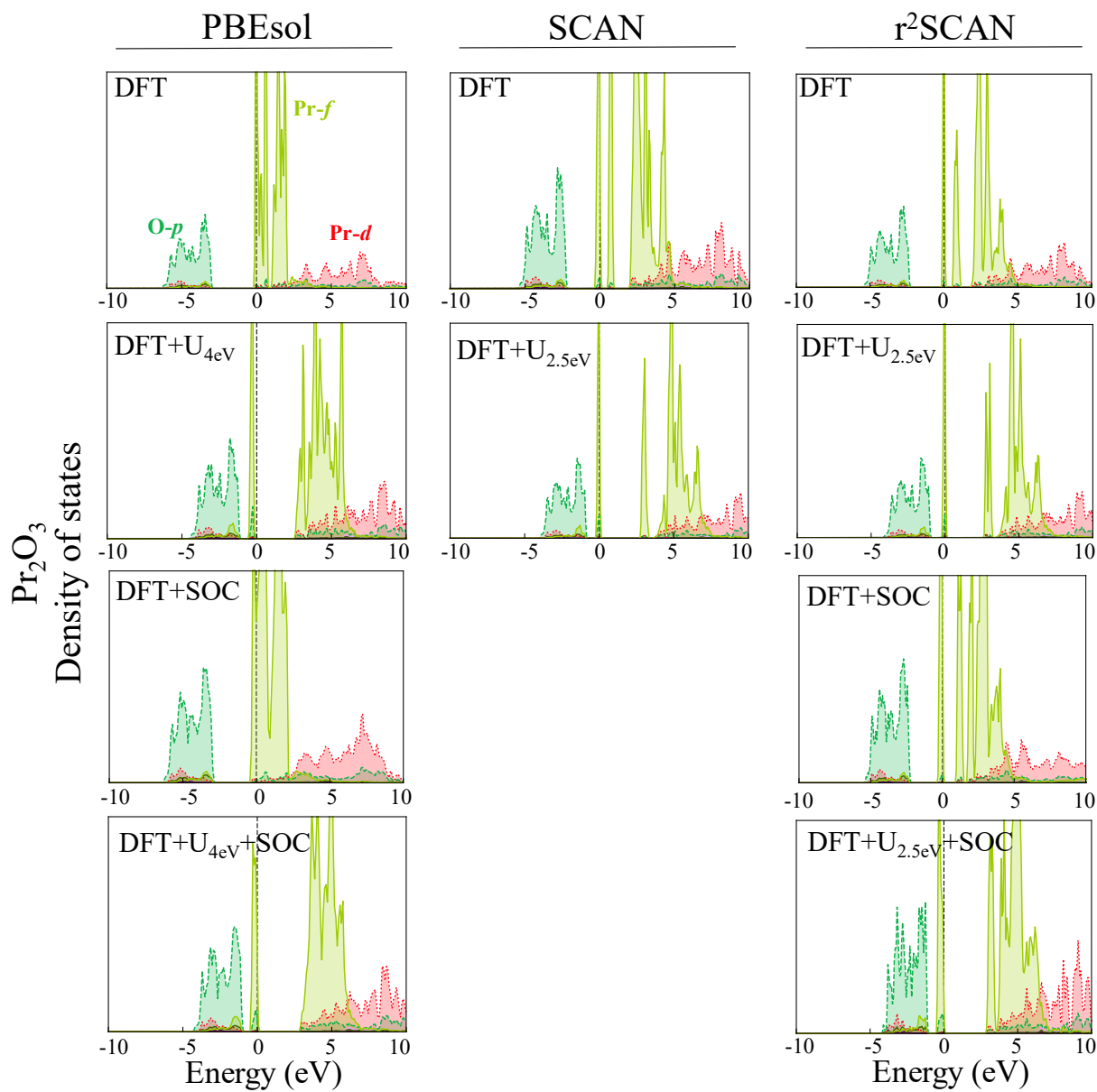

FIG. S10 DOS for bulk AFM ordered $Pr_2O_3$ calculated using DFT, DFT+U, DFT+SOC, DFT+U+SOC (DFT=PBEsol, SCAN, $r^2$SCAN).

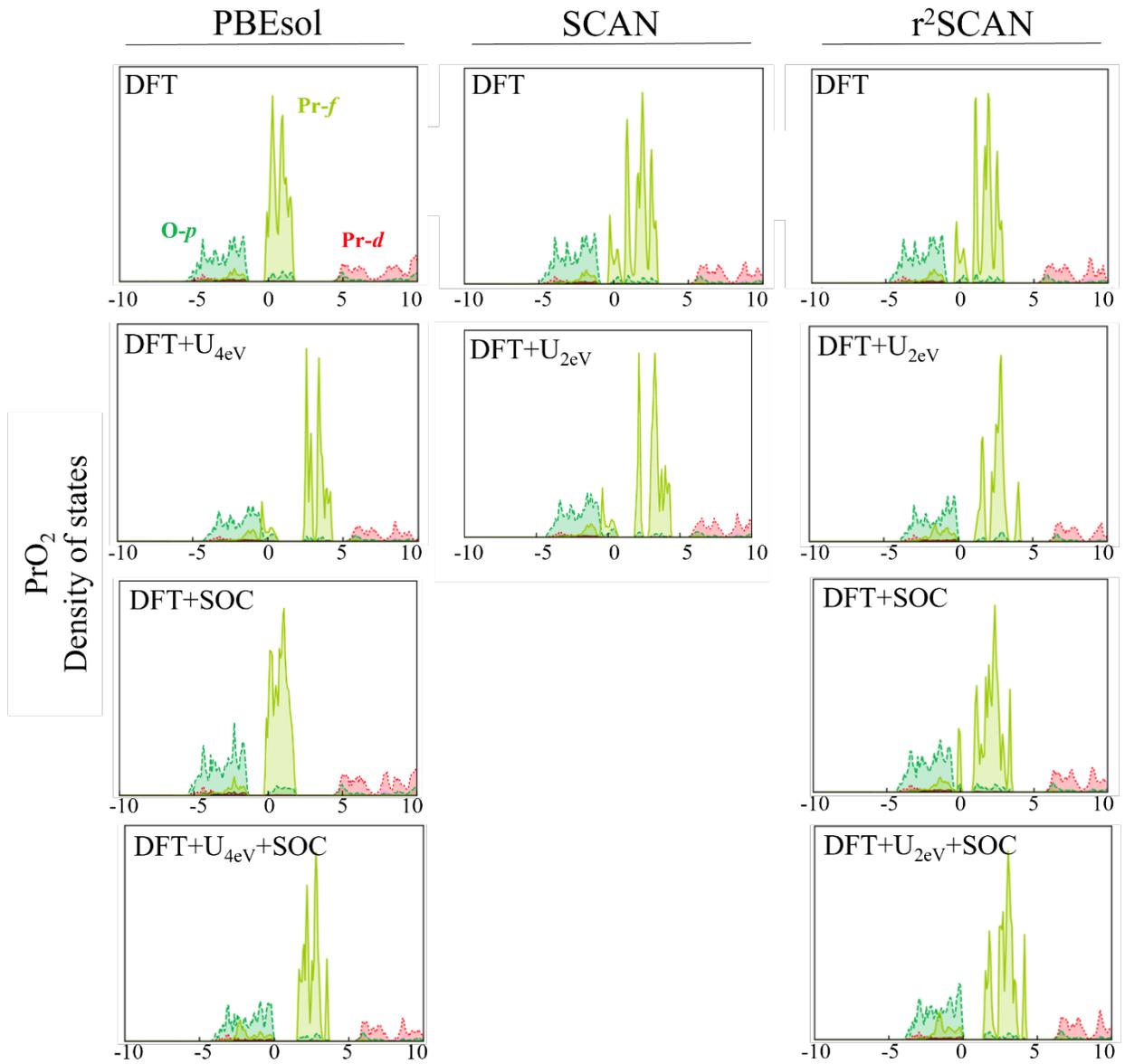

FIG. S11 DOS for bulk AFM ordered PrO$_2$ calculated using DFT, DFT+U, DFT+SOC, DFT+U+SOC (DFT=PBEsol, SCAN, r$^2$SCAN). DFT and DFT+U predict a metallic state with no band gap, except for r$^2$SCAN+U$_{2eV}$.

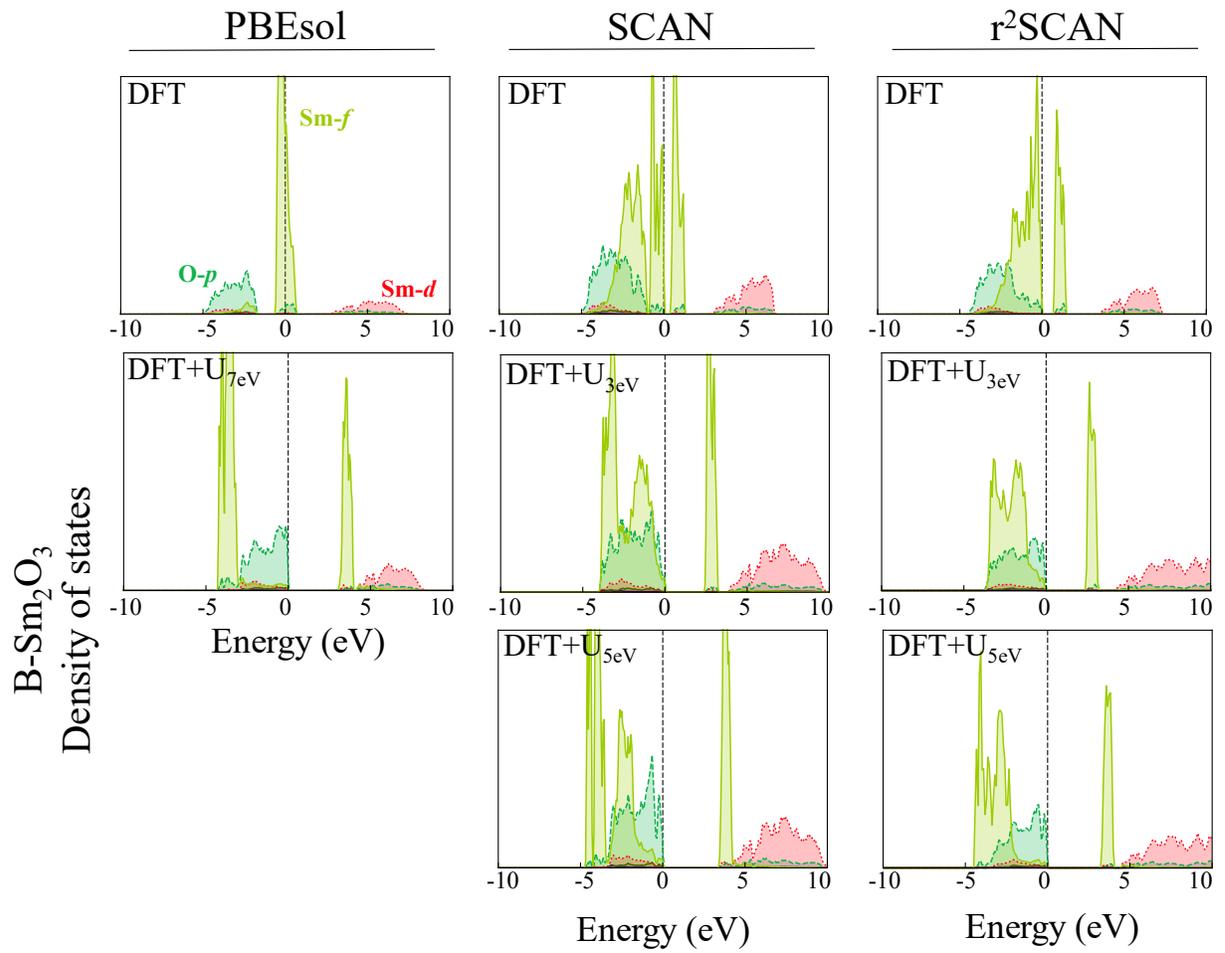

FIG. S12 DOS for bulk B-$Sm_2O_3$ calculated using DFT and DFT+U (DFT=PBEsol, SCAN, $r^2$SCAN).

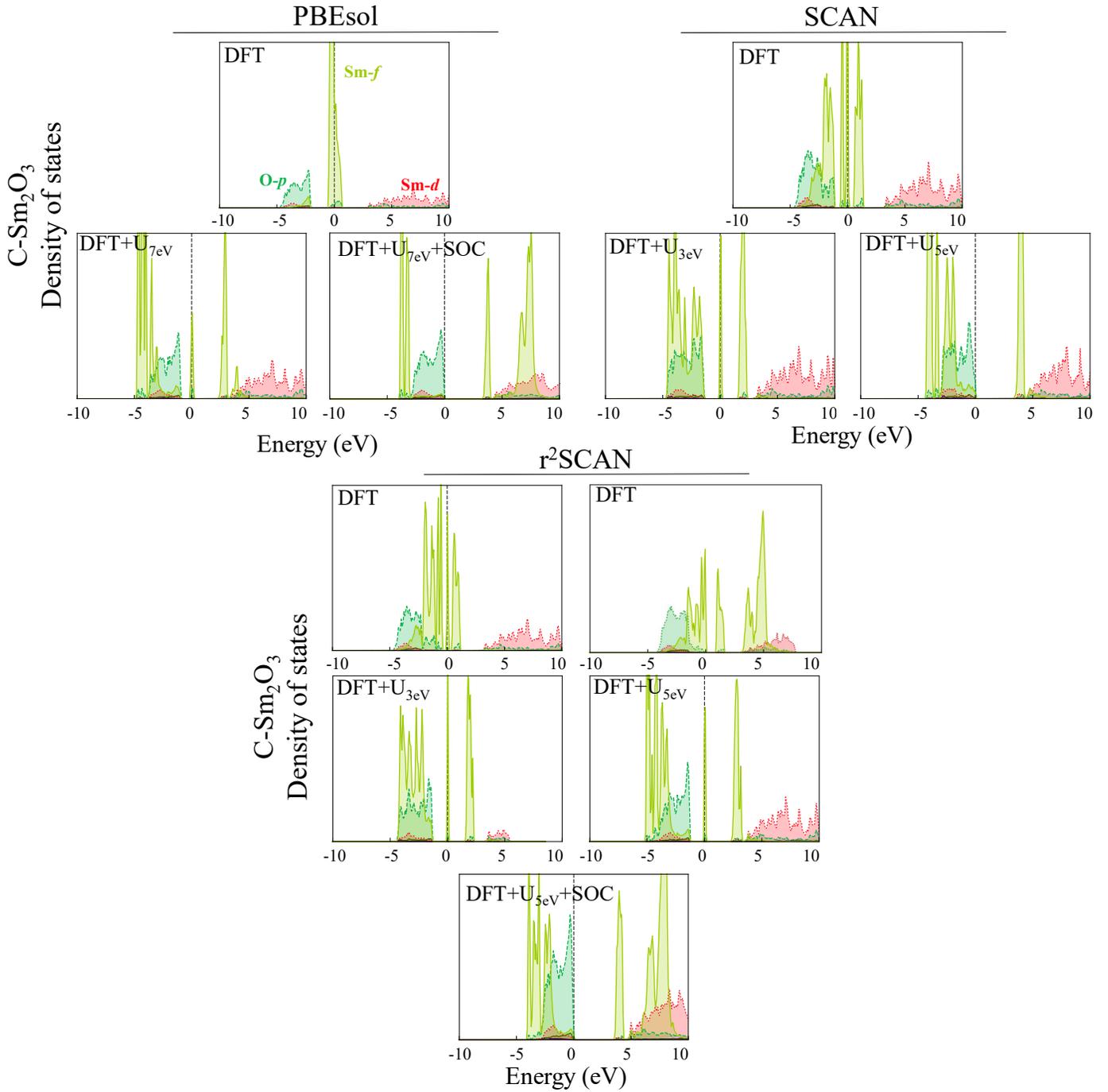

FIG. S13 DOS for bulk C-$Sm_2O_3$ calculated using DFT, DFT+$U$, DFT+SOC, DFT+$U$+SOC (DFT=PBEsol, SCAN, r$^2$SCAN).

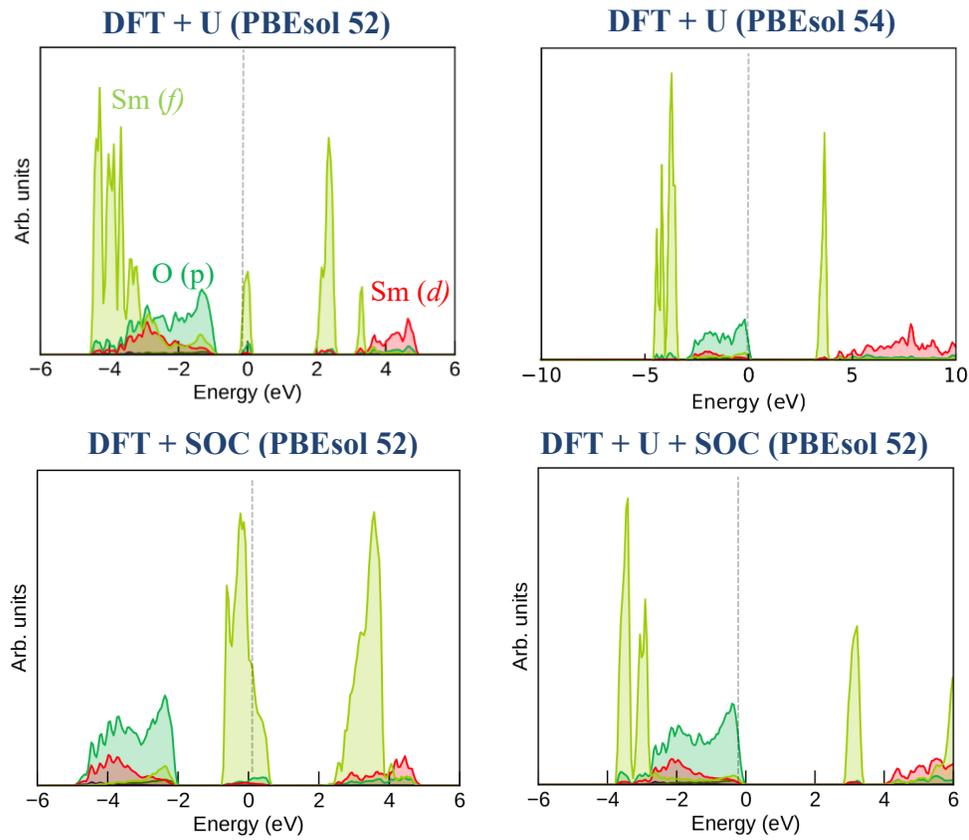

FIG. S14 DOS for bulk C-Sm$_2$O$_3$ calculated using DFT, DFT+$U$, DFT+SOC, DFT+$U$+SOC (DFT=PBEsol) in 52 and 54 PAW. DFT and DFT+SOC predict a metallic state with no band gap, except for r$^2$SCAN+U$_{2eV}$.

## Section S7. Bader charge and magnetic moment

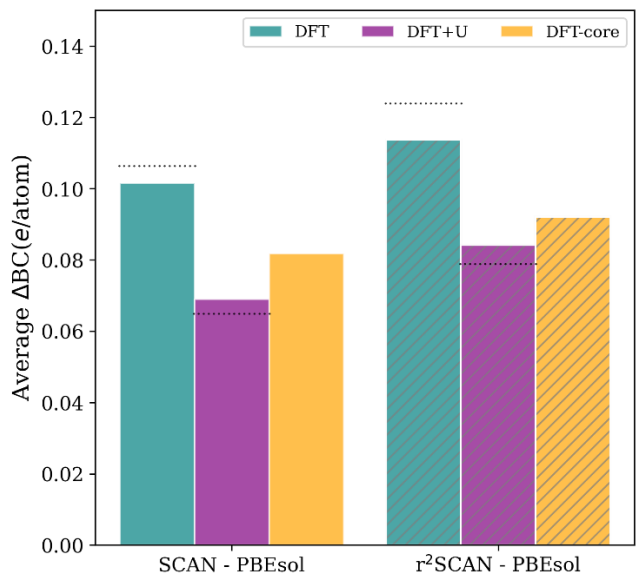

FIG S15. The average difference of Bader charge (ΔBC) between the meta-GGA functionals and PBEsol in the DFT, DFT+$U$, and DFT-core frameworks. The gray dotted lines indicate the values that contain only the *f*-band predictions for which there are *f*-core pseudopotentials, i.e., not including $Y_2O_3$, $La_2O_3$, $PrO_2$, $CeO_2$. Although the atomic charges from PBEsol+$U$ remain lower than SCAN+$U$ and r$^2$SCAN+$U$, the ΔBC for DFT+$U$ decreases [Fig. 3(b), purple bar] in comparison to the ΔBC for DFT [Fig. 3(b), light blue bar]. One exception to this was $PrO_2$ where the ΔBC increased with Hubbard $U$ due to a much larger increase in Bader charge in meta-GGA+$U$ than in PBEsol+$U$. One exception to this was $PrO_2$ where the ΔBC increased with Hubbard $U$ due to a much larger increase in Bader charge in meta-GGA+$U$ than in PBEsol+$U$.

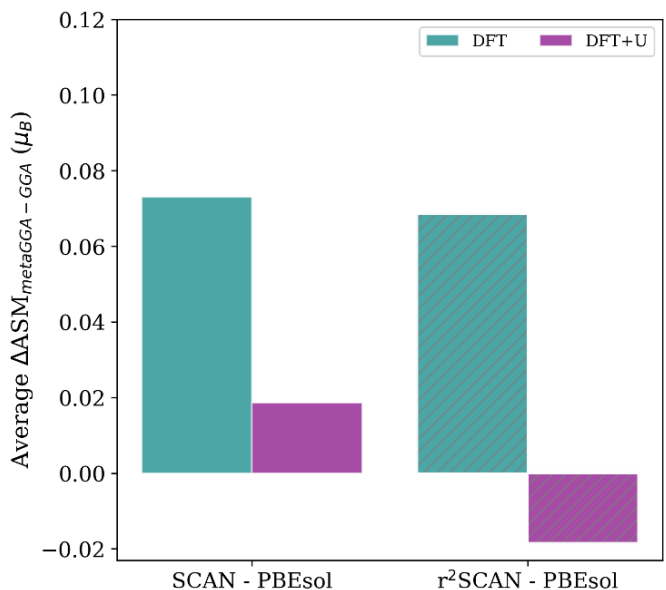

FIG S16. The average difference in atomic spin moments ΔASM between the meta-GGA functionals and PBEsol in both DFT and DFT+$U$. Positive ΔASM indicates PBEsol predicts lower magnetic moments on average compared to either SCAN or r$^2$SCAN.

TABLE S7. Cation Bader charge values (in e) for selected compounds.

| Compound | Theory | DFT | DFT+SOC | DFT+U | DFT+U+SOC | $f$ in core |
|---|---|---|---|---|---|---|
| B-Eu$_2$O$_3$ | PBEsol | 1.88 | 1.89 | 2.01 | 2.02 | 1.96 |
| | SCAN | 1.99 | | 2.08 | | 2.07 |
| | r$^2$SCAN | 2.03 | 2.03 | 2.10 | - | 2.08 |
| C-Sm$_2$O$_3$ | PBEsol | 1.89 | 1.89 | 2.00 | 2.02 | 2.01 |
| | SCAN | 2.03 | | 2.10 | | 2.09 |
| | r$^2$SCAN | 2.04 | 2.04 | 2.10 | 2.11 | 2.10 |
| A-Pr$_2$O$_3$ | PBEsol | 1.94 | 1.90 | 2.00 | 2.01 | 2.04 |
| | SCAN | 2.03 | | 2.08 | | 2.13 |
| | r$^2$SCAN | 2.05 | 2.06 | 2.08 | 2.09 | 2.13 |
| PrO$_2$ | PBEsol | 2.15 | 2.15 | 2.18 | 2.31 | 2.12 |
| | SCAN | 2.23 | | 2.25 | | 2.13 |
| | r$^2$SCAN | 2.23 | 2.31 | 2.40 | 2.37 | 2.13 |

**Section S8. Computational efficiency**

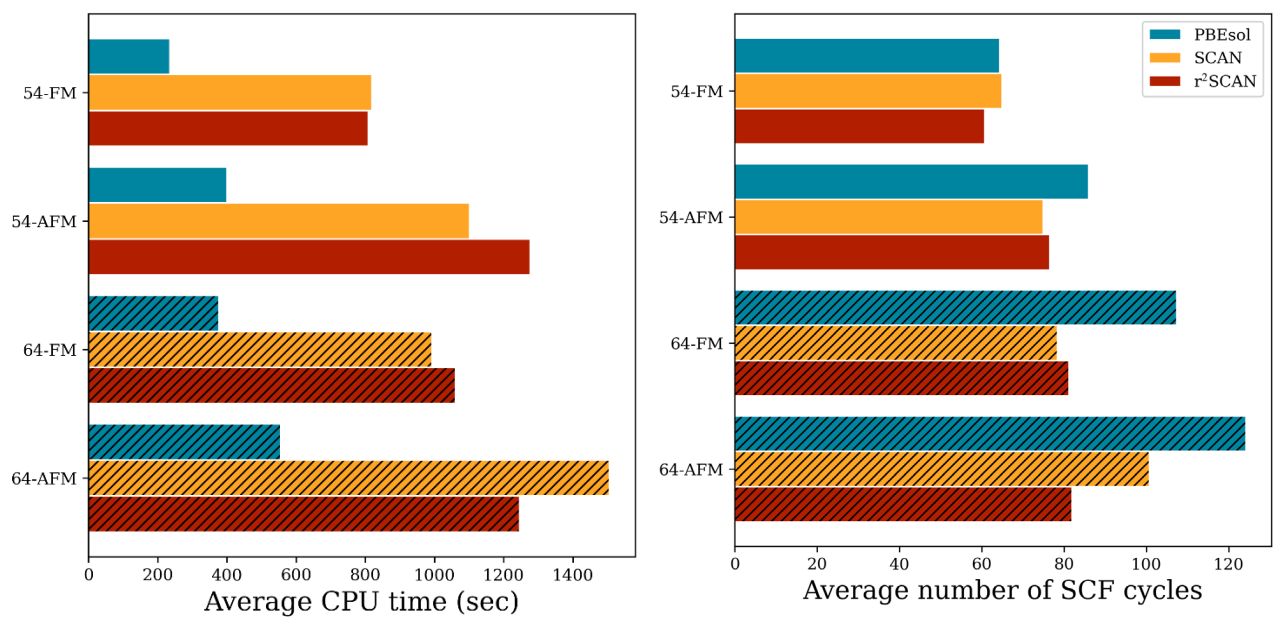

FIG S17. (a) Average computational time and (b) average number of SCF cycles of PBEsol (blue), SCAN (orange), and r$^2$SCAN (red) using the 54-and 64 versions of PAW in both FM and AFM ordering.

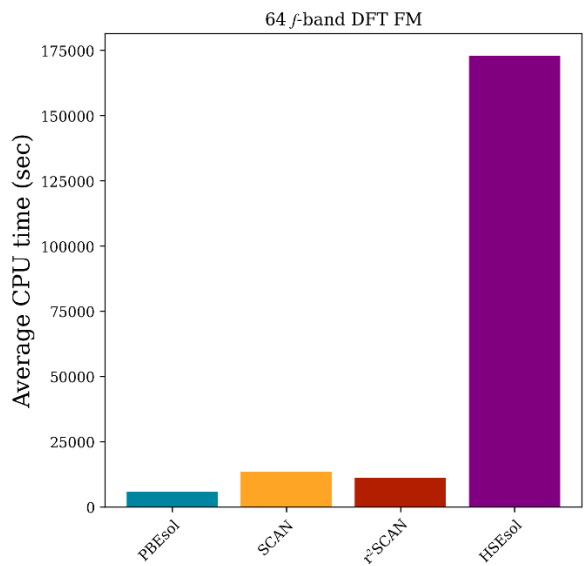

FIG S18. Average computational time (a, b) and average number of SCF cycles (c) of PBEsol (blue), SCAN (orange), r$^2$SCAN (red), and HSEsol (purple) using the 64 version of PAW in FM ordering. (b) and (c) show data only for A-Pr$_2$O$_3$ and PrO$_2$ to include comparisons for SOC calculations (diamond grid pattern). None of the hybrid calculations completed in 48 hours and only completed an average of 8 electronic steps.

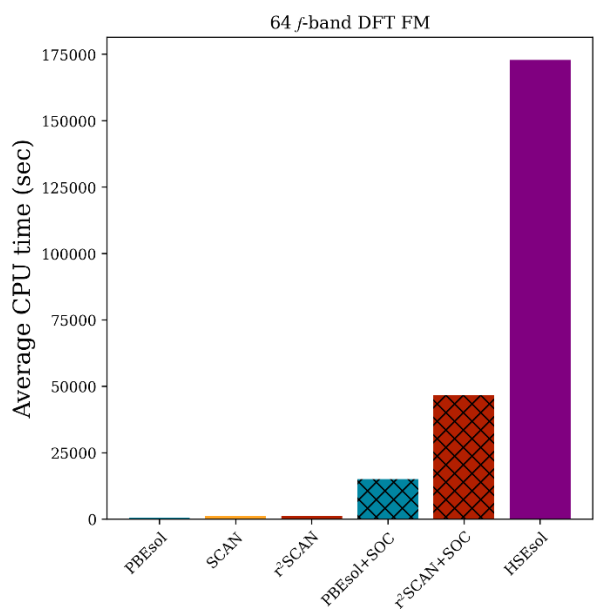

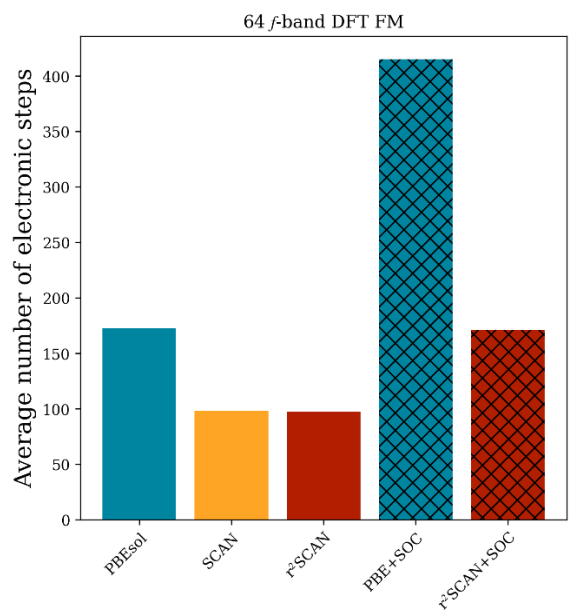

**Section S8. Notes on Computational methods**

The -DnoAugXCmeta precompiler option was not used in calculations and should not be used for the SCAN family of functionals. VASP added the (now depreciated) -DnoAugXCmeta precompiler option to address concerning behavior of the von-Weizsäcker kinetic energy density and the kinetic energy density computed from the TPSS and revTPSS orbitals. This option would only be relevant to the TPSS and revTPSS approximations since they can become unstable when the charge density is augmented. However, SCAN and its derivates (rSCAN, r$^2$SCAN, etc) are stable for the augmented density, so this option should not be used as it may negatively affect the final results.

Although yet to be formally justified, meta-GGA+SOC calculations are possible and have been previously used for formally meaningful predictions [14,15]. Including SOC via the spin-orbit coupled effective Hamiltonian allows it to be incorporated as part of routine computations in any electronic structure codes, typically applying the formalism developed by Kübler et. al. [16,17]. In the VASP code, SOC [18] and non-collinear magnetism [19] are treated with a second-variational approach and can be implemented either self-consistently [20] or non-self-consistently [21].